\documentclass[10pt,aps,prl,twocolumn,superscriptaddress]{revtex4-2}

\usepackage{amsmath}
\usepackage{amsthm}
\usepackage{amssymb}
\usepackage{times}
\usepackage{braket}
\usepackage{ulem}

\usepackage{multirow}
\usepackage{array}
\usepackage{makecell}
\usepackage{diagbox}
\usepackage{booktabs}
\usepackage{nomencl}
\usepackage{siunitx}
\usepackage{tikz}
\usepackage{placeins}
\usepackage[utf8]{inputenc}
\usepackage[dvipsnames]{xcolor}
\usepackage{hyperref}
\hypersetup{
    colorlinks=true,
    linkcolor=Maroon,
    citecolor=Maroon,
    urlcolor=Maroon,
}

\usepackage{babel}

\makeatletter
\adddialect\l@en\l@english
\makeatother

\usepackage{graphicx}
\usepackage{svg}
\graphicspath{{fig/}}

\usepackage{algorithm, algpseudocode} 
\usepackage{mathrsfs} 
\usepackage{lipsum}

\usepackage{adjustbox}

\begin{document}

\title{Colour Centre Formation in Silicon-On-Insulator for On-Chip Photonic Integration}

\author{Arnulf J. Snedker-Nielsen}
\affiliation{NNF Quantum Computing Programme, Niels Bohr Institute, University of Copenhagen, Blegdamsvej 17, 2100 Copenhagen, Denmark.}

\author{David R. Gongora}
\affiliation{University of Oslo, centre for Materials Science and Nanotechnology, PO Box 1048 Blindern, 0316, Oslo, Norway.}

\affiliation{NNF Quantum Computing Programme, Niels Bohr Institute, University of Copenhagen, Blegdamsvej 17, 2100 Copenhagen, Denmark.}

\author{Magnus L. Madsen}
\affiliation{NNF Quantum Computing Programme, Niels Bohr Institute, University of Copenhagen, Blegdamsvej 17, 2100 Copenhagen, Denmark.}

\author{Christian H. Christiansen}
\affiliation{NNF Quantum Computing Programme, Niels Bohr Institute, University of Copenhagen, Blegdamsvej 17, 2100 Copenhagen, Denmark.}

\author{Eike L. Piehorsch}
\affiliation{NNF Quantum Computing Programme, Niels Bohr Institute, University of Copenhagen, Blegdamsvej 17, 2100 Copenhagen, Denmark.}

\author{Mathias Ø. Augustesen}
\affiliation{NNF Quantum Computing Programme, Niels Bohr Institute, University of Copenhagen, Blegdamsvej 17, 2100 Copenhagen, Denmark.}

\author{Elvedin Memisevic}
\affiliation{NNF Quantum Computing Programme, Niels Bohr Institute, University of Copenhagen, Blegdamsvej 17, 2100 Copenhagen, Denmark.}

\author{Sangeeth Kallatt}
\affiliation{NNF Quantum Computing Programme, Niels Bohr Institute, University of Copenhagen, Blegdamsvej 17, 2100 Copenhagen, Denmark.}

\author{Rodrigo A. Thomas}
\affiliation{NNF Quantum Computing Programme, Niels Bohr Institute, University of Copenhagen, Blegdamsvej 17, 2100 Copenhagen, Denmark.}

\author{Mark Kamper Svendsen}
\affiliation{NNF Quantum Computing Programme, Niels Bohr Institute, University of Copenhagen, Blegdamsvej 17, 2100 Copenhagen, Denmark.}

\author{Peter Krogstrup Jeppesen}
\affiliation{NNF Quantum Computing Programme, Niels Bohr Institute, University of Copenhagen, Blegdamsvej 17, 2100 Copenhagen, Denmark.}

\author{Marianne E. Bathen}
\affiliation{University of Oslo, centre for Materials Science and Nanotechnology, PO Box 1048 Blindern, 0316, Oslo, Norway.}

\author{Lasse Vines}
\affiliation{University of Oslo, centre for Materials Science and Nanotechnology, PO Box 1048 Blindern, 0316, Oslo, Norway.}

\author{Peter Granum}
\thanks{Corresponding author}
\email{peter.granum@nbi.ku.dk}
\affiliation{NNF Quantum Computing Programme, Niels Bohr Institute, University of Copenhagen, Blegdamsvej 17, 2100 Copenhagen, Denmark.}

\author{Stefano Paesani}
\affiliation{NNF Quantum Computing Programme, Niels Bohr Institute, University of Copenhagen, Blegdamsvej 17, 2100 Copenhagen, Denmark.}

\date{
	\today
}

\begin{abstract}
Colour centres in silicon have great potential as single photon sources for quantum technologies. 
Some of them -- like the T centre -- also possess optically-active spins that enable spin-photon interfaces for generating entangled photons and multi-spin registers. 
This paper explores the generation of several types of colour centres in silicon for mass-manufacturable silicon-on-insulator quantum devices. 
We investigate how different processes in the device development affect the presence of the quantum emitters, including thermal annealing and fabrication steps for optical nanostructures. 
The study reveals coupled formation dynamics between different colour centres, identifies optimal parameters for annealing processes, and reports on the sensitivity to annealing duration and nanofabrication procedures for photonic integrated circuits.
Furthermore, we discern stable optical signals from colour centres in silicon which have not been identified before. 
\end{abstract}

\maketitle


The recent development of colour centres in silicon (SiCCs) for quantum emitters has opened up interesting prospects for the development of emitter-based photonic quantum technologies in silicon. 
Direct embedding of optically-active artificial atoms in silicon enables a promising path to scale photonic devices by leveraging the compatibility with advanced semiconductor manufacturing and integration capabilities\cite{Simmons24_Scalable,muralidharan_optimal_2016, kimble_quantum_2008}.
Over the last few years, different types of colour centres in silicon have been demonstrated to be able to emit single photons in the infrared spectrum, including the G centre (emitting near 1278~nm~\cite{Beaufils2018, redjem2020single}), T centre (1325~nm~\cite{bergeron2020silicon}), W centre (1217~nm~\cite{Baron_2022, lefaucher2023purcell}), and C centre (1570~nm~\cite{wen2025_C}), and other possible candidates e.g. I and M centres~\cite{filippatos_re-examination_2025}. 
Some of them, such as the T, M, and I centres, also possess a ground-state spin doublet that is coupled to the optical emission~\cite{higginbottom_micropucks_2022}. 
This coupling creates a spin-photon interface in the solid-state medium, enabling the generation of entanglement of multiple photons between themselves and with the system's spins - critial functionalities in quantum computing and networking applications~\cite{BarrettKok2005, Chan2025, Caleffi2024_Distributed}. \\
Technological progress with colour centres in silicon has included their integration in photonic waveguides and nanostructures to enhance photon emission~\cite{deabreu2023waveguide, lee2023high, islam2023cavity, johnston_cavity-coupled_2024, komza2024indistinguishable}, high-coherence control of electron spins coupled to the emitted photons~\cite{higginbottom_micropucks_2022}, and multi-qubit operations between long-lived electron and nuclear spins within the same colour centre~\cite{song2025longlivedentanglementspinqubitregister} as well as photon-mediated distributed operations between spins in separated T centres~\cite{PhotInc24_Distributed}.
Key to the realisation of quantum photonic devices for SiCC is the creation of the defects in silicon-on-insulator (SOI), the standard material stack for scalable photonic integration. 
Methods for generating SiCCs are commonly based on ion implantation and thermal treatment. 
Initial investigations of implantation recipes for the generation of T centres and other defects in SOI have been reported in~\cite{Buckley2020_Optimization, macquarrie_generating_2021}.
However, many aspects of the formation process of colour centres for integrated quantum devices in SOI remain undetermined, such as the coupled formation dynamics between the optical defects, optimal process parameters for the different colour centres, and the effects of nano-device fabrication procedures. \\
Here, we study how different processes in the development of silicon-based integrated quantum photonic devices affect colour centre formation. 
We examine the effects of a wide range of implantation and thermal annealing parameters, including both temperature and annealing time, as well as individual fabrication steps for the integration with photonic nanostructures. 
We simultaneously monitor various types of SiCCs, including T, I, M, C, G, and W colour centres, to investigate how their formation processes vary and how they correlate. 
Detailed parameter scans reveal dynamics in their formation, enabling the observation of activation, suppression, and revival mechanisms when varying annealing temperatures. 
We determine optimised annealing parameters and fabrication procedures for the formation of colour centres in SOI, some of which we find to be significantly different from values previously reported in the literature~\cite{macquarrie_generating_2021}.
Furthermore, we identify new emission lines in SOI that are stable under variations of the formation parameters and likely originate from novel colour centres. 
The suite of SiCC formation processes analysed and optimised in this study can be of valuable guidance in the development of future emitter-based devices for scalable quantum photonic technologies in silicon and for future studies of formation mechanisms of defects in silicon in general. 

\begin{figure*}[t]
    \centering
    \includegraphics[width=1\linewidth]{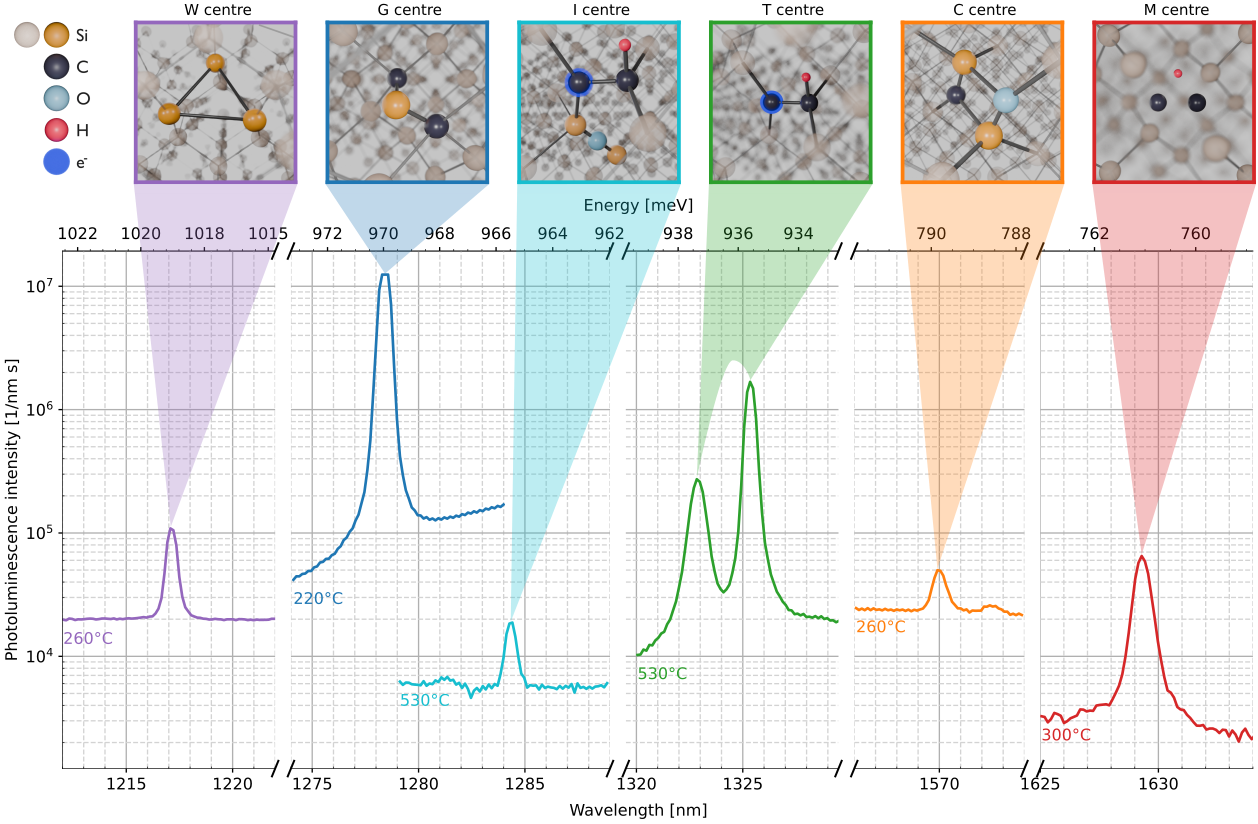}
    \caption{Characteristic PL spectra of prominent colour centres demonstrating their presence in the 1200 to 1600 nm range after annealing at different temperatures. The atomic structures for the defects associated with each peak are also shown in the insets~\cite{filippatos_re-examination_2025}, except for the M centre, for which the atomic structure is still undetermined. For each defect, we show the PL spectra with the highest photoluminescence signal amongst those obtained with the different annealing parameters we tested. The associated annealing temperature is reported below each of the peaks.
    }
    \label{fig:Annealed_spectra}
\end{figure*}

\begin{figure*}[t]
    \centering
    \includegraphics[width=1\linewidth]{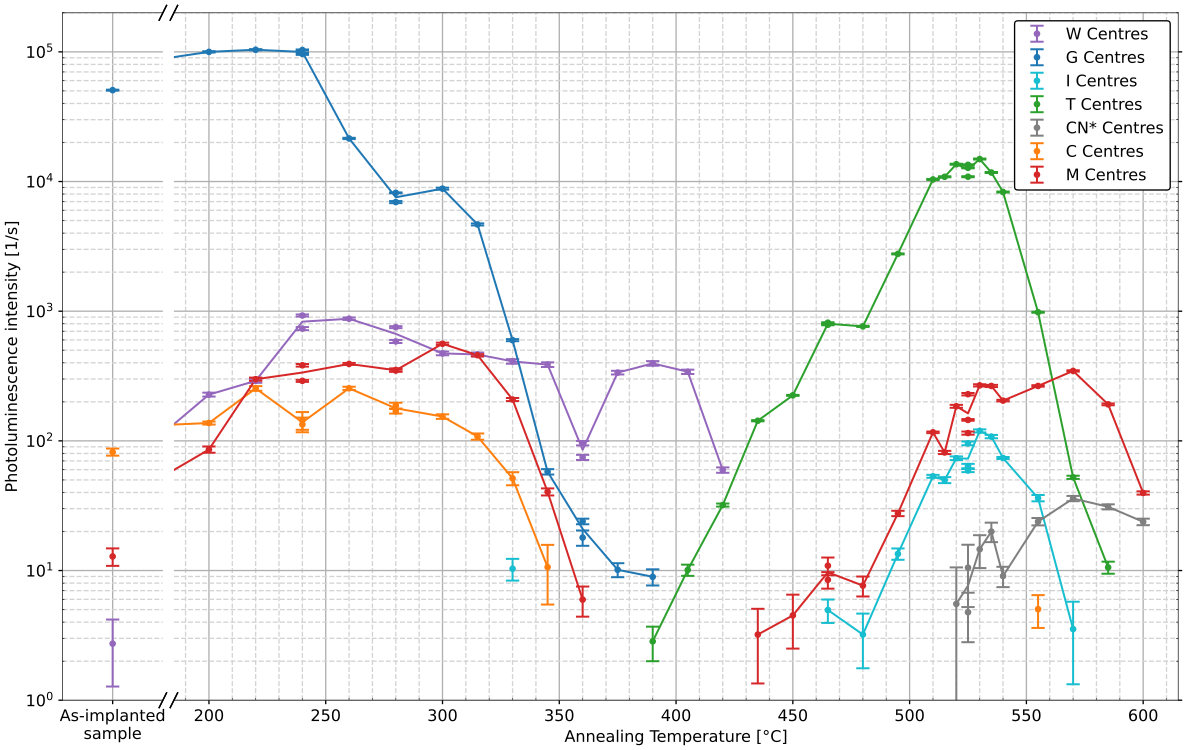}
    \caption{Photoluminescence signal from several optically active colour centres as a function of the activation annealing temperature. The intensity of the signal is extracted as the area of a Gaussian on a second order polynomial background fitted to a window around the peak. Error bars denote $\pm 1\sigma$ statistical uncertainty.
    }
    \label{fig:centres_vs_AnnealTemp}
\end{figure*}


\section{Formation procedures and experiments}
\label{sec:method}

The material platform used for the fabrication is commercially available 220nm SOI wafers ($\geq 3000$ $\Omega$ cm resistivity).
Implantation of carbon and hydrogen ions on the SOI wafers is performed either through commercial implantation services (Ion Beam Services) or at the NorFab ion implanting facilities at the University of Oslo.
The ion implantation follows standard recipes, with a 1:1 fluence between the two species~\cite{song2025longlivedentanglementspinqubitregister, macquarrie_generating_2021}.
Carbon is implanted first, followed by a \SI{1000}{\celsius} curing anneal for $20$~s to repair the damage in the Si crystal structure induced by the carbon implantation, then followed by hydrogen implantation.
All measurements reported in the main text are done on SOI with a $^{12}$C implantation fluence of $7\times 10^{13}$cm$^{-2}$.
A detailed description of the material and measurements on a wafer with a lower implantation fluence ($1\times 10^{13}$cm$^{-2}$ of $^{13}$C) can be found in Appendix~\ref{sec:supp:detailesWafer}. However, the low density wafer shows qualitatively similar results to the main wafer. 
After ion implantation, the SOI wafers are processed for the colour centre's activation and nano device fabrication steps.
The activation process consists of low-temperature annealing, for which we investigate temperatures ranging from 200 to \SI{600}{\celsius} and annealing times ranging from 30 to 600 seconds.
Moreover, we explore different procedures, such as the inclusion of a boiling step in deionised water preceding the activation, process gas flows ranging from 200 to 5000 sccm, and both optical and contact measurements for thermal control.
The fabrication steps investigated are a broadly representative set of typical processes used to fabricate silicon photonic devices with e-beam lithography (EBL). 
We study the effect of each step on the SiCC formation by producing sequences of silicon chips, with each chip obtained by interrupting the device fabrication after a targeted step.
Immersion in an oxygen plasma for resist removal (ashing) is investigated for a range of durations from 30 to 600 seconds, as well as compared with remote ashing and different orderings of nanofabrication steps in the larger processing scheme. \\
To examine the effect of annealing on the distribution of the implanted C and H ions, we performed secondary ion mass spectroscopy (SIMS) measurements, both pre- and post-annealing on samples implanted with a $5\times 10^{14}$cm$^{-2}$ C$^{12}$ fluence. 
The measured data, which is reported in Appendix~\ref{sec:supp:detailedSIMS}, shows a carbon atom distribution centred around the middle of the membrane at a depth of 110~nm, and the hydrogen distribution is centred around a depth of 120 nm.
The measurements also show that while the vertical distribution of C was not visibly affected by the subsequent thermal annealing, the hydrogen atoms gradually diffuse through the silicon at higher annealing temperatures. This is expected, as the small size of hydrogen means that it diffuses easily in silicon~\cite{pearton1987hydrogen}, whereas carbon is much more immobile, relying on more intricate mechanism for diffusion in silicon~\cite{scholz1999contribution}.
Characterisation of the colour centres is performed by collecting the photoluminescence (PL) spectra from the processed samples.
The PL spectra are recorded using a short-wave infrared (SWIR) cryogenic setup operating at 4.8~K. 
Three lasers at 405, 532, and 852~nm are used as excitation sources to study defect densities at different implantation depths, and the PL signal is acquired using a spectrograph coupled to an InGaAs array detector. 
The recorded PL signals span the range from 1040 to 1660~nm, limited by the detection range of the InGaAs detector in the spectrometer. More details on the PL measurement setup and data analysis procedures are reported in Appendix~\ref{sec:supp:detailedSetup} and Appendix~\ref{sec:supp:PLAnalysis}, respectively.
PL signals from colour centres are observed when the cryostat temperature is brought below 40~K. Our standard operating temperature for all data reported in the main manuscript is 4.5~K (see Appendix~\ref{sec:supp:tempRedShift} for measurements done at higher temperatures). 
This allows us to study and identify a wide range of optically active colour centres in silicon, some of which have, to the best of our knowledge, an unknown atomic structure. 
A list of some of the colour centres identified is given in Table~\ref{tab:silicon_defects}, with a more extensive list of detected defect signatures provided in Appendix~\ref{sec:supp:detailedListSiCC}. 
Note that the exact transition energies for each defect may vary slightly with temperature, local magnetic and electric fields, nearby impurities, surface proximity, and the local strain environment~\cite{Clear_2024}. 
%


\section{Study of the annealing process}
\label{sec:temperature_annealing_scan}
To examine the effects of the activation annealing step on colour centre formation, we annealed a sequence of samples for 3 minutes in a nitrogen atmosphere at temperatures between 200 and \SI{600}{\celsius}.
The annealing temperature steps are smaller than \SI{20}{\celsius} across the whole range to obtain a detailed picture of the SiCC behaviour. 
We also performed experiments that included a boiling step before the activation anneal, as is sometimes described in the literature~\cite{macquarrie_generating_2021, higginbottom_micropucks_2022,johnston_cavity-coupled_2024}, although we find this to have a negligible effect on the final photoluminescence, as shown in Appendix~\ref{sec:supp:qualitative_variations}.
In Fig.~\ref {fig:Annealed_spectra} we show some of the PL peaks observed below the bulk silicon band-edge emission energy (1060~nm at 4~K) using a 405 nm excitation laser at approximately $30~\text{W}/\text{cm}^2$.  
The peaks in Fig~\ref{fig:Annealed_spectra} correspond to the W, G, I, T, C, and M colour centre emissions.
We also show their generally accepted atomic structure~\cite{Baron_2022, Beaufils_2018, Gower_1997, Nakamura_1995, Safonov1996}.

\begin{table}[t!]
    \centering
    \setlength{\tabcolsep}{4pt}
    \caption{Identified colour centres and their corresponding zero-phonon line (ZPL) energies and wavelengths. $\text{T}_{\text{II},\text{max}}$ denotes the annealing temperature at which the maximum integrated PL intensity was observed. Entries labelled with ``(?)'' correspond to defects with either unknown or speculative atomic structure. Reported errors are $\pm1\sigma$ statistical uncertainties.}
    \begin{tabular}{lccccc}
        \toprule
        Centre & ZPL  & ZPL & $\text{T}_{\text{II},\text{max}}$ & Atomic \\
         & [nm] & [eV] & [°C] & structure \\
        \midrule
        W & 1217.17 $\pm$ 0.02 & 1.02  & 240     & I\textsubscript{3} (tri-interstitial Si) \\
        G & 1278.6  $\pm$ 0.05 & 0.970 & 200     & C\textsubscript{s}-Si\textsubscript{i}-C\textsubscript{s} \\
        I & 1284.36 $\pm$ 0.04 & 0.965 & 530     & C-C-H(O) (?) \\
        T & 1325.40 $\pm$ 0.04 & 0.935 & 530     & (C-C)\textsubscript{s}-H\textsubscript{i} \\
        C & 1569.99 $\pm$ 0.05 & 0.790 & 240     & C\textsubscript{i}-O\textsubscript{i} \\
        M & 1629.33 $\pm$ 0.07 & 0.761 & 300/570 & C-C-H (?) \\
        \bottomrule
    \end{tabular}
    \label{tab:silicon_defects}
\end{table}

The W, G, C, and M centres are already detectable in the as-implanted sample, before any activation annealing treatment. 
All these complexes are carbon-related except the W centre, which is a self-interstitial Si aggregation defect with an emission energy at 1218~nm. 
It is attributed to a tri-interstitial (I\textsubscript{3}) silicon cluster, although its precise atomic configuration remains under discussion~\cite{Baron_2022, Gennetidis_2024,COOMER1999505}. 
W centre formation is driven by the implantation-induced lattice damage, which displaces Si atoms and generates both Si self-interstitials (Si\textsubscript{i}) and Si vacancies (V\textsubscript{Si}). 
Since Si\textsubscript{i} is a highly mobile defect, it can aggregate under suitable annealing conditions to form the I\textsubscript{3} clusters. 
Under the implantation and annealing conditions used in this study, we observed that the W centre population is stable up until \SI{420}{\celsius}, where its optical signal becomes unmeasurably low, indicating a stark drop in W centre density and a partial recovery of the Si crystalline structure\cite{Davies06}. \\
%

%
Amongst the carbon-related defects observed in the as-implanted samples, the G centre exhibits the strongest PL signal. 
This defect consists of two substitutional carbon atoms, C\textsubscript{s}, connected by Si\textsubscript{i} and emitting at 1278~nm. 
Its concentration shows no significant change between the as-implanted sample and the activation annealing treatments up to \SI{240}{\celsius}, indicating that the conditions required for its formation are only satisfied during the initial carbon implantation and the subsequent curing anneal. 
The carbon implantation initially generates C\textsubscript{i} defects, but also produces Si\textsubscript{i} and Si V\textsubscript{Si} through the displacement of Si atoms from their lattice sites. 
During annealing, a fraction of the C\textsubscript{i} atoms are incorporated into the lattice, forming carbon substitutional (C\textsubscript{s}), and also forming  C\textsubscript{s}C\textsubscript{i} pairs. 
One specific configuration of these di-carbon defects has been reported to promote G centre formation~\cite{Deak_2023}. 
Notably, the decrease in the relative formation of the G centre starts at annealing temperatures above \SI{240}{\celsius} and becomes undetectable at around \SI{400}{\celsius}. 
%

%
In the case of the C centre -- a C\textsubscript{i}-O\textsubscript{i} complex with an emission line at 1569~nm~\cite{Magnea_1984} -- its maximum population is observed at around \SI{260}{\celsius}. Any further annealing progressively reduces its population, which becomes undetectable at annealing temperatures above \SI{340}{\celsius}.
Annealing above \SI{260}{\celsius} may convert the C centres into optically inactive carbon-oxygen-related defects, such as C\textsubscript{i}-O\textsubscript{i}(Si\textsubscript{i})~\cite{Backlund_2007, Backlund_2008}. 
Alternatively, the thermal treatments may promote the dissociation of the C centre complexes, releasing C\textsubscript{i} and O\textsubscript{i} atoms that can subsequently participate in the formation of other defect structures.
%

The M centre is a carbon-hydrogen-based defect emitting at 1629~nm. 
Its formation is proposed to occur via the reaction  (C\textsubscript{i}-H\textsubscript{i})+C\textsubscript{s}$\rightarrow$ C-C-H, although this mechanism remains under debate~\cite{Filippatos_2025}. 
The low-temperature annealing reveals two different temperature windows suitable for M centre formation. 
In the first window, ranging from as-implanted to \SI{360}{\celsius}, the M centre population exhibits a plateau between 220 and \SI{320}{\celsius} and becomes undetectable above \SI{360}{\celsius}. 
We then observe a revival of M centres in a second window, for annealing temperatures above \SI{435}{\celsius}. M centres reappear with increasing temperature and reach their local maximum density at \SI{570}{\celsius}, before decreasing again when approaching \SI{600}{\celsius}. \\
All carbon-based defects observed in the as-implanted sample reach their minimum relative emission at around \SI{400}{\celsius}, leaving a temperature window between roughly \SI{360}{\celsius} and \SI{420}{\celsius} in which the optical activity is negligible. 
This behaviour suggests that, within this range, carbon -- predominantly in the form of highly mobile C\textsubscript{i} -- begins to form complexes with other ions, giving rise to optically inactive carbon-related defects and reducing the populations of the G, C, and M centres. 
One of these intermediate carbon-related dark complexes, stable between \SI{360}{\celsius} and \SI{420}{\celsius}, may act as precursors that promote the subsequent formation of the T centre. 
On the other hand, unlike the other carbon-based centres, the M centre is not observed only in the temperature interval between \SI{360}{\celsius} and \SI{435}{\celsius}, suggesting that the precursor complex responsible for M centre formation remains stable before the low-temperature treatment and within this intermediate temperature range.
In Fig.~\ref{fig:centres_vs_AnnealTemp} we also observe that the G centres disappear around \SI{400}{\celsius}, which is where the T, M and I centres appear, indicating that their (de)formation dynamics are linked. 
At \SI{600}{\celsius}, the high temperatures seemingly cause all bonds between the defects to break, and optical signals from all the tracked colour centres decrease significantly or are already undetectable. \\
%

At annealing temperatures around \SI{390}{\celsius}, the T centre becomes observable. 
This centre is a well-studied defect whose atomic structure is generally described as a carbon-pair complex occupying a silicon lattice site (C-C)\textsubscript{s} and bonded to a H\textsubscript{i} atom. 
As shown in Fig.~\ref{fig:Annealed_spectra}, the T centre's ZPL emission is located at 1326~nm. 
Due to the internal strain associated with the colour centre within the silicon crystal, the exciton undergoes a splitting into two distinct doublet energy levels, denoted $\text{TX}_0$ and $\text{TX}_1$, separated by approximately 1.75~meV~\cite{Clear_2024}. 
The annealing reveals that T centre formation reaches a maximum density at around \SI{525}{\celsius} and becomes undetectable above \SI{570}{\celsius}. 
Previous studies suggest that T centre formation relies on reactions between  C\textsubscript{s} defects and C\textsubscript{i}-H\textsubscript{i} complexes generated during the ion implantation~\cite{Safonov1996,SAFONOV1999}. 
This mechanism is further supported by a recent DFT study \cite{Dhaliah2022}, which identified this pathway as energetically feasible with a thermodynamic barrier of 1.75~eV, although alternative formation pathways were also proposed. 
An activation barrier of 1.75~eV is consistent with the 390 to \SI{550}{\celsius} temperature window observed for G centre out-annealing and T centre formation in Fig.~\ref{fig:centres_vs_AnnealTemp}.
Previous works find an optimal annealing temperature for T centre formation in SOI of $\sim$\SI{450}{\celsius}~\cite{macquarrie_generating_2021}. Consistently, a  DFT study predicted a similar temperature range \cite{Dhaliah2022}, showing that optimal T centre formation requires a hydrogen chemical potential of $-0.92$~eV, corresponding to \SI{450}{\celsius} at a hydrogen partial pressure of $10^{-6}$~bar. 
This contradicts our experimental observation, where the maximum T centre population occurs at $\sim$\SI{525}{\celsius}. 
To possibly explain this discrepancy, we studied the influence of changing the thermal processing (RTP) settings used for the annealing, but we found that different settings cannot explain the observed discrepancy in optimal annealing temperature (further details on this are presented in Appendix~\ref{sec:supp:qualitative_variations}). \\
%

The I centre is the carbon-based defect that forms in the 450 to \SI{550}{\celsius} range. 
Its proposed atomic structure is similar to that of the T centre, but with an oxygen atom located in proximity to the defect, forming a C–C–H(O) complex with an emission energy of 1285~nm~\cite{Gower_1997}. 
As shown in Fig.~\ref{fig:centres_vs_AnnealTemp}, the formation ratio of the I centre follows the same temperature-dependent trend as that of the T centre, and like the T centre, the I centre reaches its maximum relative emission at \SI{530}{\celsius} and becomes undetectable above \SI{570}{\celsius}, suggesting that I and T centres may have similar formation mechanisms. 
Notably, the I centre emission becomes detectable at around \SI{460}{\celsius} -- approximately \SI{90}{\celsius} higher than for the T centre -- which coincides with the temperature range at which O\textsubscript{i} becomes mobile in silicon \cite{Lightowlers_1994}. 
This oxygen may be released during the dissociation of C centres. 
Our observations further supports the interpretation that the I centre corresponds to a T centre–like complex with the additional involvement of oxygen atoms. 

\begin{figure}
    \centering
    \includegraphics[width=1\linewidth]{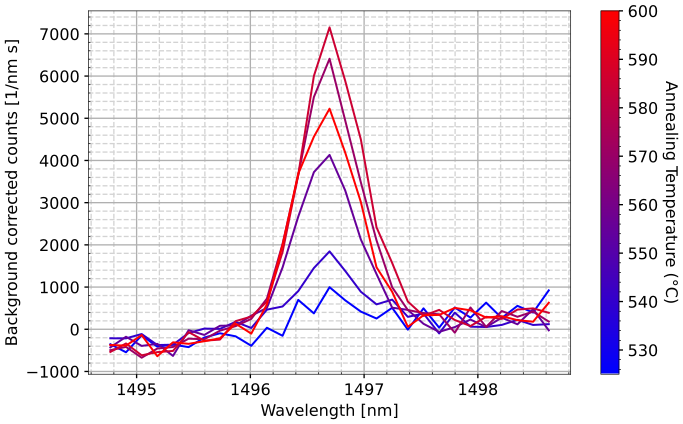}
    \caption{Spectra with an unidentified photoluminesce peak centred around 1496.7~nm, which we call CN* due to association with the centre proposed in Ref.~\cite{nangoi2025cncomplexalternativet}. 
    The signal arises only after annealing at \SI{540}{\celsius}, but increases in brightness up to \SI{570}{\celsius}. 
    The signal from the material response has been subtracted from the signal for ease of visualisation. 
    }. 
    \label{fig:unidentified_peak}
\end{figure}


\section{Formation of novel colour centres}
In addition to the colour centres discussed so far, which are already described in the literature, we have also identified clear and stable PL signals which have not been identified before.
We highlight PL peaks that are stable across a range of annealing temperatures of at least \SI{50}{\celsius}. 
A list of all identified new peaks is reported in Appendix~\ref{sec:supp:detailedListSiCC}.
One of these signals is located in the low-loss telecom S-band and appears at annealing temperatures above \SI{520}{\celsius}, as shown in Fig.~\ref{fig:unidentified_peak}.
The annealing temperature dependence of the peak is also reported in Fig.~\ref{fig:centres_vs_AnnealTemp}.
The signal can be observed to grow at temperatures above \SI{520}{\celsius}, with a maximum at \SI{570}{\celsius}, indicating that it is robust to temperatures where most other SiCCs are breaking apart. 
The central wavelength of the peak is $1496.66~\pm~0.05$~nm (statistical uncertainty), corresponding to 0.828 eV. 
This measured emission energy is close to predictions for a new carbon-nitrogen-based defect that has been recently theoretically proposed in Ref.~\cite{nangoi2025cncomplexalternativet}, which estimates a ZPL line at 1497.4~nm  from first-principle calculations.
Furthermore, the temperature range in which this peak appears matches the temperatures where other carbon-nitrogen-related lines have been seen to form, indicating mobility of nitrogen in the lattice \cite{dornen_complexing_1988,dornen_nitrogen-carbon_1986,dornen_set_1987,dornen_vibrational_1986,dornen_nitrogen-carbon_1986}.
We, therefore, tentatively label this line CN* due to these indications of it matching with properties of the CN centre recently proposed in Ref.~\cite{nangoi2025cncomplexalternativet}. 
Theoretical calculations estimate that this new centre,  the (CN)$_\text{Si}$ centre,  has many of the same attributes as the T centre, including comparable lifetime, Debye-Waller factor, and a spin-triplet ground state, and might thus prove a useful alternative for building photonic quantum technologies due to the more stable nature of the nitrogen bound in the lattice and the predicted less energetically favourable decomposition reactions. 
This is consistent with nitrogen diffusing into the lattice and combining with substitutional carbon-defects to form the joint-substitutional defect.


\section{Effects of annealing time}
\label{sec:duration_annealing_scan}

\begin{figure}
    \centering
    \includegraphics[width=1\linewidth]{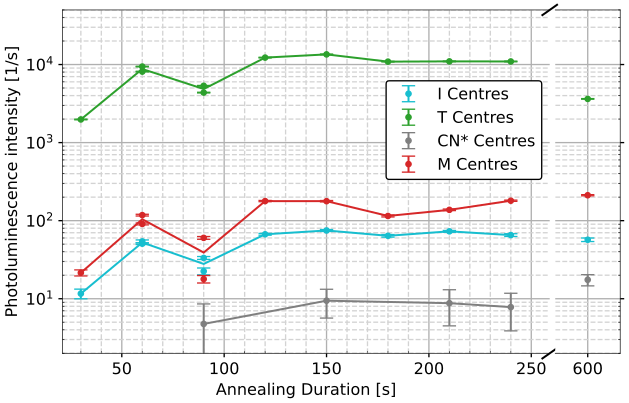}
    \caption{Photoluminescence signal from different Colour centres for samples annealed at \SI{525}{\celsius}, shown for a variable duration of the activation anneal. Error bars denote $\pm 1\sigma$ statistical uncertainty.}
    \label{fig:duration_scan_peaks}
\end{figure}

In addition to the temperature dependence, we examine the impact of the duration of the activation annealing step, since previous studies have suggested kinetic inhibition as an important mechanism for avoiding the breakdown of T centres, which requires control of the process duration on the order of the T centre breakdown rate~\cite{Dhaliah22}.
To investigate this aspect, we annealed samples at 525 $^\circ$C for a variable duration from 30 to 600 seconds. 
Photoluminescence results are reported in figure~\ref{fig:duration_scan_peaks}, which shows the intensity of the peaks associated with the T, I, and M centres as a function of annealing time. 
Beyond 120 seconds, the intensities prove to be stable for time scales of a few minutes. 
This allows a better comparison between our and other works~\cite{macquarrie_generating_2021,song2025longlivedentanglementspinqubitregister,johnston_cavity-coupled_2024}, as differences in temperature rise times etc. of different equipment can be ignored. 
The reduction in T centre density after 10 minutes of annealing indicates that the annealing duration can be used as a fine-tuning knob for the T centre density, as an alternative to changing the annealing temperature.


\section{Effects of fabrication steps}

\begin{figure*}
    \centering
    \includegraphics[width=\linewidth]{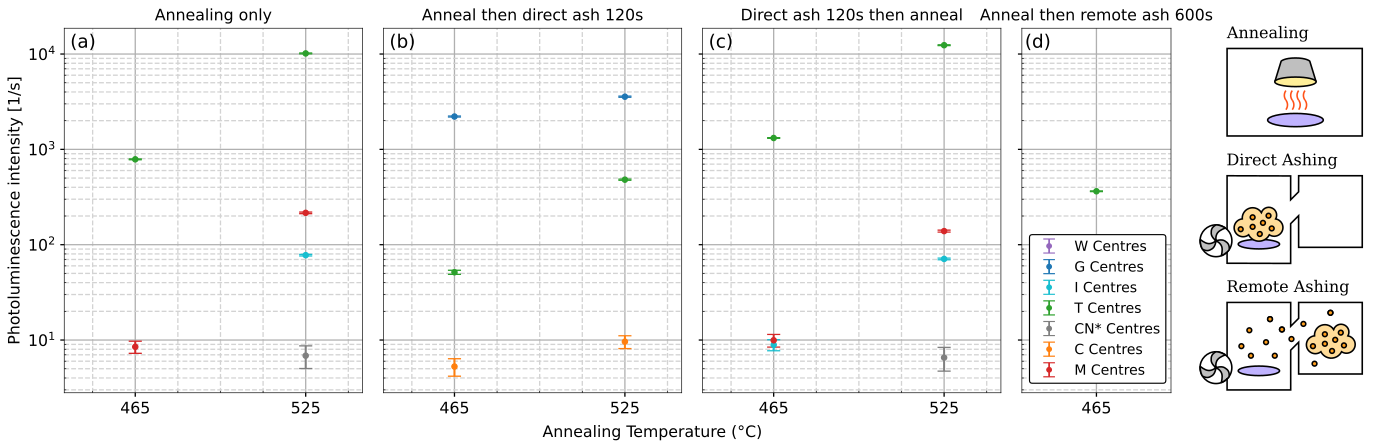}
    \caption{Effects of ashing on colour centre formation. \textbf{(a)} Reference samples only undergoing the activation anneal similar to samples presented in Figure~\ref{fig:Annealed_spectra}. \textbf{(b)} The effect of ashing on annealed centres indicating a reduction in the T, I, and M centre densities but an increase for the G and C centre. \textbf{(c)} The effect of ashing before annealing, showing similar densities as the reference samples. \textbf{(d)} The effect of ashing with a remote plasma. Despite the longer ashing period, the effect on emitter density is negligible, meaning this presents an alternative to direct ashing in cases where ashing must be done after activating colour centres but where it is undesirable to harm existing centres. Error bars denote $\pm 1\sigma$ statistical uncertainty.
    }
    \label{fig:ashing_sweep}
\end{figure*}

The integration of colour centres in photonic integrated circuits (PICs) is critical for their potential applications in quantum technologies. 
We investigated the stability of colour centres during an electron beam lithography (EBL) process to examine their integration with CMOS compatible fabrication methods. 
We truncated the fab process after spin coating, e-beam exposure, silicon etch, and dry stripping the resist respectively. After interrupting the fabrication, we used a long wet strip in solvent as the most gentle way of removing layers of resist on the surface before measuring the PL spectra of the samples.
While we do not see significant effects from most processes, we do see a drop in the colour centre densities during the ashing process used to strip the resist after fabrication. The impact of ashing was confirmed by ashing the samples without exposing them to other processing, yet a decrease in colour centre densities still occurs. 
The PL measurements for all fabrication steps are shown in Appendix~\ref{sec:supp:fabrication} as well as the results for different durations and methods of ashing.\\
Further experiments, including switching the order of the annealing and nanofabrication processes, found that the effect of the ashing process is distinctly different from simply heating the sample, since it also reintroduces  centre and C centres previously annealed out of the sample.
Figure~\ref{fig:ashing_sweep} shows peak intensities as a function of different cleaning procedures on samples annealed at 465 or \SI{525}{\celsius}, showing both the reduction in T centre luminescence and the reactivation of G centres which probably results from the decomposition of optically inactive defects. 
The ashing tool used (Tergeo Plasma Cleaner) supports both a direct and remote plasma source, meaning plasma is respectively generated in the primary chamber with the sample or in a secondary remote chamber and indirectly introduced to the sample.
Performing the entire PIC fabrication before the final annealing step proved to yield similar T centre densities to those of samples not exposed to PIC fabrication. We hence suggest this order of fabrication steps.
Further nanofabrication steps that must be applied to the material stack, i.e. for quantum computing, should not heat the sample to more than $\sim$\SI{400}{\celsius} for extended periods to avoid affecting the density of T centres as suggested by Fig.~\ref{fig:centres_vs_AnnealTemp}. 
Additionally, we have shown that high-intensity plasma cleaning also decreases T centre density, but this effect can be avoided by either delaying the final annealing step until after ashing, as shown in Fig.~\ref{fig:ashing_sweep}c, or by using gentle remote ashing, as seen in Fig.~\ref{fig:ashing_sweep}d. 
Remote ashing might prove particularly useful in cases where colour centres are present in the substrate as-grown and need to be preserved without an annealing step, such as in Ref. \cite{Epitaxial_SiCC_Aberl24}.
%


\section{Conclusion}
We have reported a comprehensive analysis of the formation of colour centres in silicon through annealing processes and of how the centres are affected by device nano-fabrication methods. 
We have studied coupled formation dynamics between various types of colour centres during the annealing process, with the results confirming that annealing temperature serves a key role in determining which colour centres are activated.
For the T centre, a prominent candidate for scalable photonic quantum technologies in silicon, we find an optimal activation annealing temperature of $525\pm$\SI{10}{\celsius}, which is substantially different from the values reported in previous studies.
Furthermore, we observe that some processes often reported in the formation of T centres, such as boiling samples in deionised water, do not result in significant effects on the detected optical signal.
For other analysed SiCCs that represent competitive alternatives to the T centre, such as the I, C, and M centres,  our experiments represent the first analysis of optimal formation parameters. 
Furthermore, our comprehensive exploration of annealing parameters has enabled us to discern photoluminescence signals from defects that have not been identified before. 
In particular, we observe a clear and stable optical signal in the S-band that matches the predicted ZPL of the so far unobserved CN colour centre, a new type of SiCC that has been theoretically proposed only very recently and represents a promising alternative to the T centre for quantum computing and networking.
These results can guide the exploration of photonic quantum devices with novel types of quantum emitters in silicon. 
%


\section{Acknowledgements}

The authors acknowledge funding support from the NNF Quantum Computing Programme and the Research Council of Norway (research grants No. 325573 and 295864). S.P. acknowledges funding from VILLUM FONDEN (MapQP, No. VIL60743), the European Research Council (ERC StG ASPEQT, No. 101221875), and Danmarks Innovationsfond research grant No. 4356-00009B (HyperTenQ). 

\section{Competing Interests}
The authors declare no competing interests.

\section{Data Availability}
An online repository with the datasets reported in this study is made available at Ref.~\cite{online_repo}.



\bibliography{library}


\newpage 
\onecolumngrid
\clearpage

\appendix

\setcounter{secnumdepth}{2}

\setcounter{figure}{0}
\setcounter{table}{0}

\renewcommand{\thefigure}{\Alph{section}\arabic{figure}}
\renewcommand{\thetable}{\Alph{section}\arabic{table}}


\section{Wafer and Implantation}
\label{sec:supp:detailesWafer}
\begin{figure}[h]
    \centering
    \includegraphics[width=\linewidth]{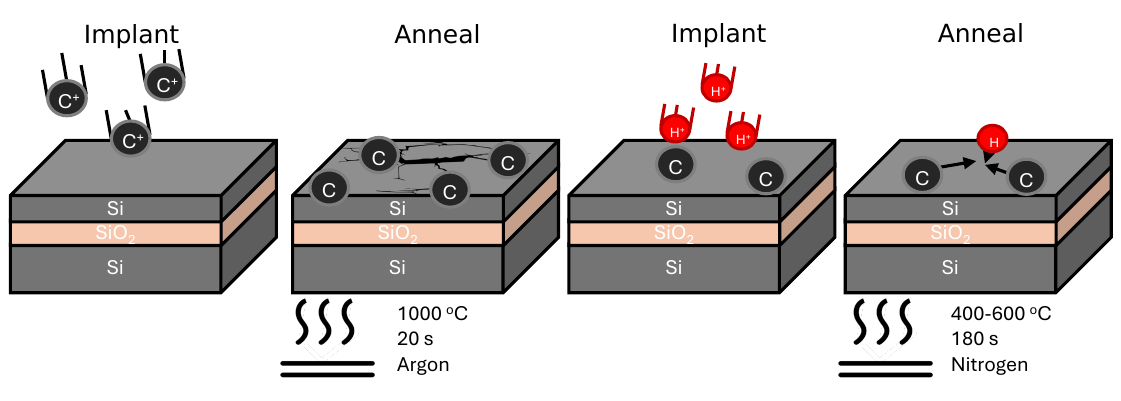}
    \caption{Implantation and annealing process for generating colour centres on SOI. The SOI wafer is first implanted with carbon which is then annealed at 1000 C for 20 s in an argon atmosphere. The sample is then implanted with hydrogen, which is finally annealed to form the hydrogen-carbon colour centres.}
    \label{fig:Implantation_workflow}
\end{figure}

The SOI wafers were bought from Shin-Etsu and consist of 220 nm of Si on 3 \textmu m on SiO$_2$ on 725 \textmu m Si with a resistivity of $\geq 3000$ $\Omega$ cm. Ion Beam Services implanted carbon and protons at 38 and 9 keV respectively with a stoichiometric ratio of 1:1. The primary samples were implanted with $7\times 10^{13}$cm$^{-2}$ of $^{12}$C, while another set of samples was implanted with $1\times10^{13}$cm$^{-2}$ of $^{13}$C. The implantation angle was perpendicular to the sample surface (no tilt). Carbon was implanted first, followed by a 20 s anneal at \SI{1000}{\celsius} in an argon atmosphere to heal the crystal damage caused by the implantation as illustrated in Figure~\ref{fig:Implantation_workflow}. Protons were implanted as a final step. From the wafers, smaller samples were cleaved and annealed at lower temperatures to activate the colour centres. The activation annealing step was done in an RTP (AccuThermo AW 610) using a lidded graphite susceptor and a nitrogen flow rate of 5000 sccm for the duration of the anneal, unless otherwise noted (see Appendix \ref{sec:supp:qualitative_variations}). \\ \\

\begin{figure}
    \centering
    \includegraphics[width=\linewidth]{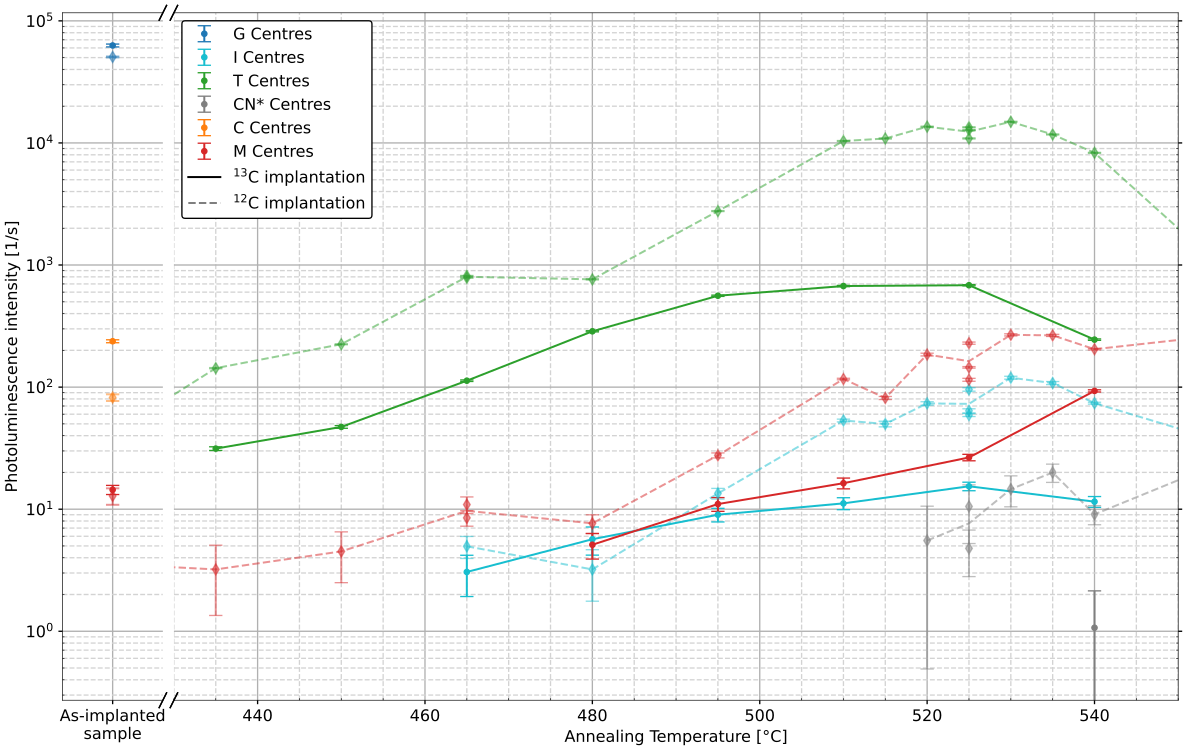}
    \caption{Colour centre densities for different temperatures of the activation annealing step for the sample implanted with $^{13}$C at a fluence of $1\times10^{13}$cm$^{-2}$ (solid line) compared to the results for the same annealing parameters presented in Fig.~\ref{fig:centres_vs_AnnealTemp} (dashed line). 
    %
    %
    \label{fig:low_fluence_anneal}}
\end{figure}

The results in the main text are all derived from the $^{12}$C-implanted wafer, but similar results can be seen from samples with the lower fluence $^{13}$C implantation. As shown in Fig.~\ref{fig:low_fluence_anneal}, the optimal temperature for T centre generation is shifted slightly lower (this is not evident for the I or M centres) and the overall density of all defects is lower post-annealing, replicating previous results for a similar implantation in Ref.~\cite{macquarrie_generating_2021}.

\section{Secondary Ion Mass Spectroscopy}
\label{sec:supp:detailedSIMS}
To determine the vertical distribution of C and H through the SOI, a set of samples was measured using secondary ion mass spectrometry (SIMS) with a Cameca IMS 7f magnetic sector instrument. A beam of Cs$^+$ 15 keV ions was rastered over an area of 200 × 200 \textmu m$^2$ with a beam current of 50 nA, and data was collected for the central part of the sputtered crater (diameter of 62 \textmu m). The samples exposed to SIMS were implanted with H and C to a fluence of $5 \times 10^{14}$ cm$^{-2}$ -- a slightly higher fluence than that investigated in the main paper -- to provide a hydrogen concentration substantially above the detection limit of SIMS. The samples were annealed at \SI{420}{\celsius} and \SI{525}{\celsius} respectively. \\

\begin{figure}[h]
    \centering
    \includegraphics[width=1\linewidth]{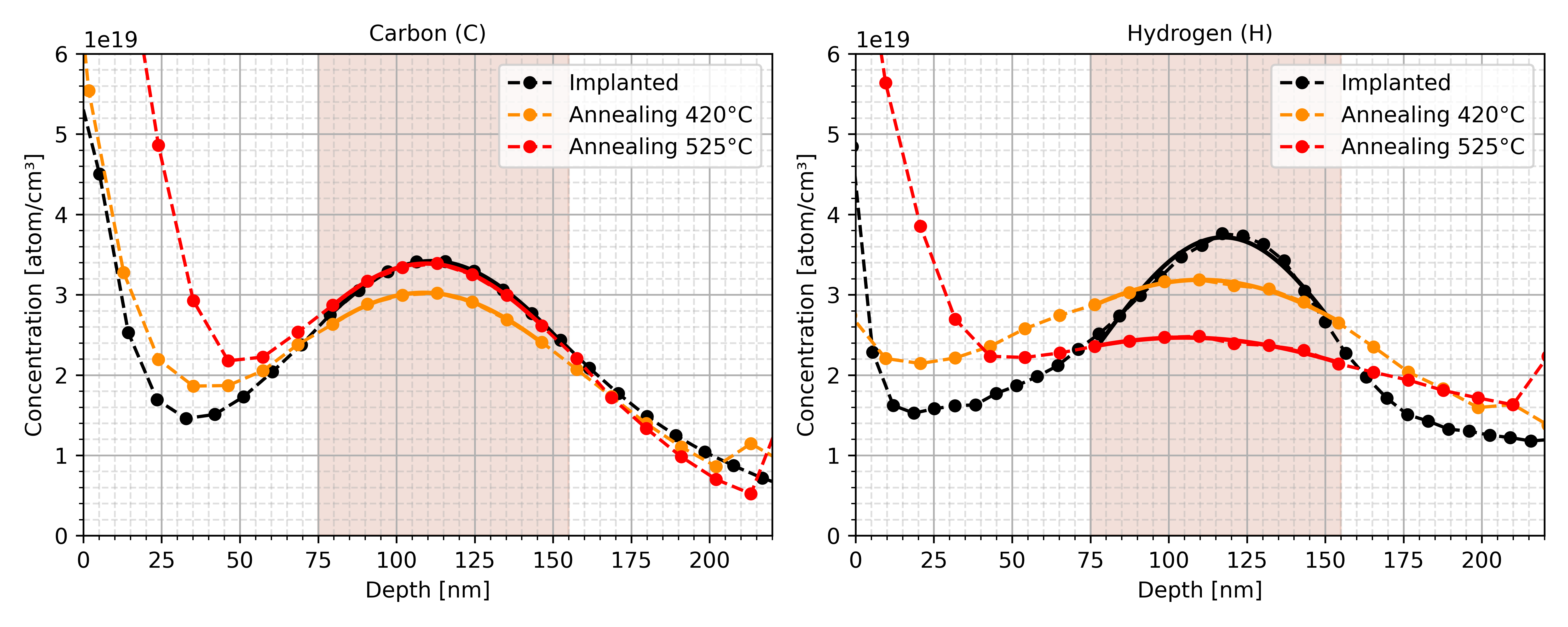}
    \caption{SIMS measurements showing carbon and hydrogen concentrations as a function of depth for as-implanted and annealed samples at \SI{420}{\celsius} and \SI{525}{\celsius}. The dashed line connects data points while the solid lines are Gaussian fits to the data within the highlighted region. See Tab. \ref{tab:SIMS_fit} for the parameters of the fits. The elevated concentration of hydrogen and carbon from 0 to ~25 nm originates from surface effects and is not representative of the bulk composition of the SOI membrane.}
    \label{fig:sims}
\end{figure}

\begin{table}[h]
\label{tab:SIMS_fit}
\setlength{\tabcolsep}{18pt}
\caption{Gaussian fit parameters of the SIMS data to the model $A\exp(-(x-x_0)^2/(2\sigma^2))$. The hydrogen FWHM is seen to increase with annealing activation temperature indicating a greater mobility of the hydrogen as opposed to the carbon under the activation annealing.}
\begin{tabular}{lccc}
\hline
Sample & Amplitude [atom/cm$^3$] & Centre [nm] & FWHM [nm] \\
\hline
C Implanted           & $(3.42 \pm 0.06) \times 10^{19}$ & $111.3 \pm 0.15$ & $116.3 \pm 0.78$ \\
C Annealing 420°C     & $(3.02 \pm 0.06) \times 10^{19}$ & $108.7 \pm 0.21$ & $131.0 \pm 1.31$ \\
C Annealing 525°C     & $(3.39 \pm 0.06) \times 10^{19}$ & $109.4 \pm 0.18$ & $121.3 \pm 1.04$ \\
H Implanted           & $(3.71 \pm 0.38) \times 10^{19}$ & $117.5 \pm 0.68$ & $99.5 \pm 3.19$ \\
H Annealing 420°C     & $(3.19 \pm 0.14) \times 10^{19}$ & $110.2 \pm 0.74$ & $174.5 \pm 5.30$ \\
H Annealing 525°C     & $(2.47 \pm 0.14) \times 10^{19}$ & $104.8 \pm 1.89$ & $224.0 \pm 14.1$ \\
\hline
\end{tabular}
\end{table}

Figure \ref{fig:sims} shows the distribution of the depth of the carbon (Fig.~\ref{fig:sims} (a)) and hydrogen (Fig.~\ref{fig:sims} (b)) atoms before and after annealing. The distribution is seen to be centred around 120 nm, as opposed to the uniform distribution suggested in \cite{komza2024indistinguishable}. The measurements also reveal that hydrogen diffuses during annealing, where a gradual reduction around the target region is observed, but a significant amount of hydrogen still remains in the region after \SI{525}{\celsius} annealing. On the other hand, the C profile is similar for all the measured samples, indicating no significant diffusion.

\section{Photoluminescence Methods}
\label{sec:supp:detailedSetup}

\subsection{Optical Setup}
The photoluminescence (PL) measurements were performed at ~4.8 K using a closed-cycle helium cryostat. Optical excitation was provided by continuous-wave lasers with wavelengths of 405 nm (Oxxius LaserBoxx LBX-405-300-CSB-PP), 532 nm (Toptica DL 100), and 852 nm (Millenia eV), each operating at a power of 10 mW with a beam radius of 100 \textmu m. This power proved low enough to not saturate the emitters (see section \ref{sec:supp:PLAnalysis}) but high enough to allow a clear signal. For most measurements, the 405 nm laser was used, as it deposits more power in the SiCC region - see section \ref{sec:supp:powerDeposition}. The laser beams were focused onto the sample surface at an incident angle of approximately 30°, resulting in an excitation intensity of $\sim30$~W/cm² for all sources.\\
The PL signal was collected in a backscattering geometry using a 10x apochromatic near-infrared microscope objective (Mitutoyo, 0.26 NA, 31 mm working distance). The emitted luminescence was passed through a long-pass filter (LP800) to suppress scattered excitation light and was analysed using an imaging spectrometer (Horiba iHR550), equipped with 300 and 1200 grooves/mm gratings. The spectrometer was coupled to an InGaAs detector array (Andor DU491A), providing a spectral resolution of $\sim 0.14$~nm with the 300 gr/mm grating and $\sim 0.04$~nm with the 1200 gr/mm grating. The sample and beam spot were imaged before measurement by inserting a 90(R)/10(T) beam splitter in the collection path. A dichroic mirror with transmission above 950 nm was inserted in the collection path when using the 852 nm laser to remove the laser in the collection window.  
\begin{figure}
    \centering
    \includegraphics[width=0.8\linewidth]{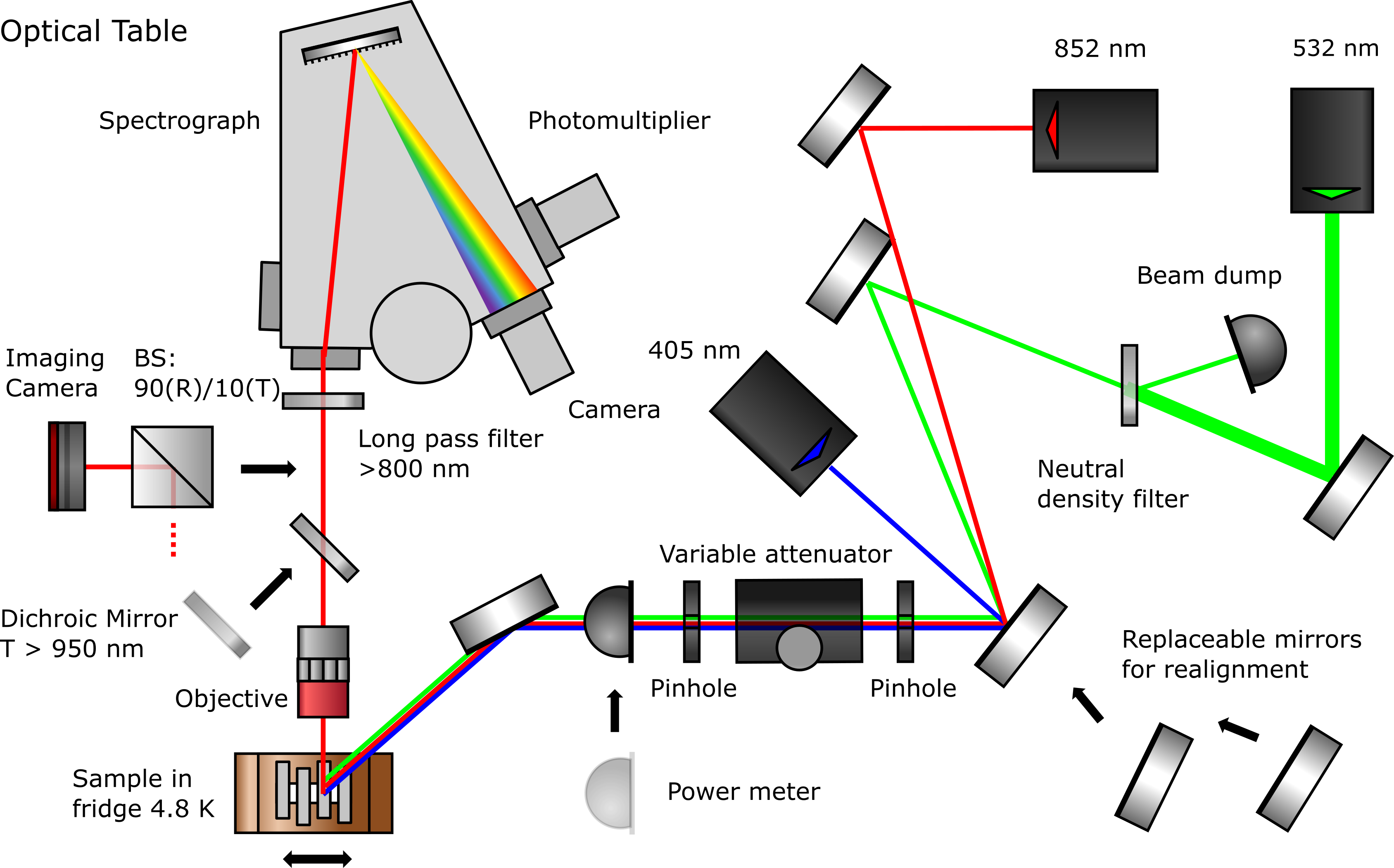}
    \caption{Sketch of the PL setup. Three different lasers at 405, 532 and 852 nm were used. Two pin holes were used for initial alignment. Replaceable mirrors with magnetic feet were used for easy and reliable alignment of the three lasers. A variable attenuator and a power meter were used to control the power of the beam. A final mirror was used to guide the beam into the cryostat at a $30^\circ$ incidence angle onto the cryostat window and samples. The cryostat was operated at 4.8 K. A 10x magnification objective was used to collect the PL signal which was passed through a long pass filter $>$800 nm to filter out the lasers. For the 852 nm laser a dichroic mirror T$>$950 nm was inserted into the collection path. An image of the sample was acquired by inserting a (90 R, 10 T) beam splitter. Finally, the signal was sent into a spectrograph optimized using a lens to resolve different colour centre signals from the sample. }
    \label{fig:setup}
\end{figure}

\subsection{Saturation \& Optical Excitation Power}
\begin{figure}[h]
    \centering
    \includegraphics[width=\linewidth]{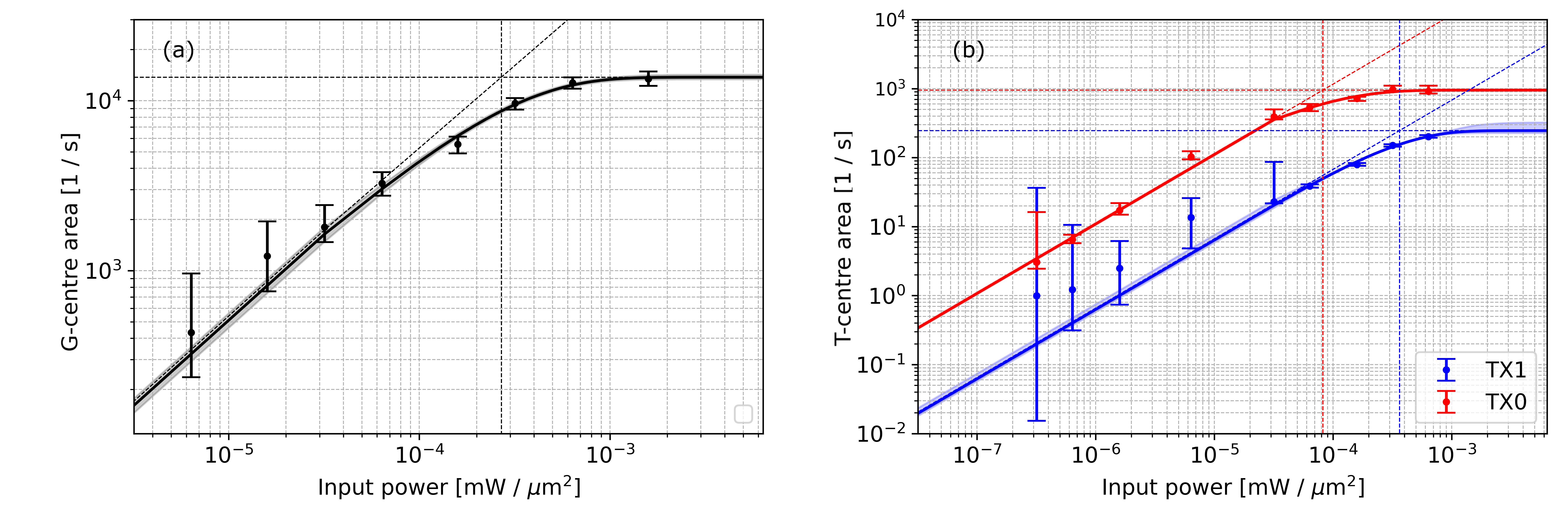}
    \caption{The power dependence of the G-centre and T-centre brightness as a function of the input power. The data has been fitted to Eq. \ref{eq:pl_power_dependence}. \textbf{(a)} The G-centre PL emission saturates at 8.5 (7.8, 9.3) mW, corresponding to 2.7 (2.5, 3.0) $\times 10^{-4}$mW / \textmu m$^2$, with a saturation brightness of 2.8 (2.7, 3.2)$\times10^6$, and a model order of 0.969 (0.911, 0.995), which is close to unity as expected from a first-order process. \textbf{(b)} The TX0 line saturates at 2.6 (2.3, 2.8) mW, with a saturation brightness of 19.1 (18.7, 19.3)$\times10^4$, and an order of 0.80 (0.76, 0.83). The TX1 line saturates at 11 (9, 19) mW, with a saturation brightness of 4.9 (4.5, 6.5)$\times10^4$ , and an order of 1.0 (0.9, 1.1), indicating that the TX1 line was not fully saturated.}
    \label{fig:power_dependce}
\end{figure}

The probability of a single emitter emitting a photon is proportional to the probability that the emitter is excited. Photoluminescence  signals can be fitted using a simple power law $I(P)\propto P^n$, where $n$ is typically $1\leq n \leq2$ depending on the underlying absorptive and radiative processes. We describe the event of an emitter being excited as a stochastic process. The probability of exciting a single emitter $k$ times within a given period follows a Poisson distribution
\begin{equation}
    f(k,\lambda) = \frac{\lambda^k e^{-\lambda}}{k!}
\end{equation}
where $\lambda$ is the mean number of excitations within a given period. The mean number of excitations must be proportional to $(P/P_\text{sat})^n$, where $P_\text{sat}$ denotes the saturation power and ensures that $\lambda$ is unit-less, and $n$ accounts for non-linear excitation processes such as two-photon absorption ($n=2$). The probability of the emitter not getting excited within a given period then becomes
\begin{equation}
    f\left(0,(P/P_\text{sat})^n\right) = e^{-(P/P_\text{sat})^n}
\end{equation}
The intensity of photoluminescence from a single emitter is proportional to the probability of the emitter being excited at least once within a certain time frame. Hence the emission probability distribution for a single emitter is given as
\begin{equation}
    1 -   f\left(0,(P/P_\text{sat})^n\right) = 1- e^{-(P/P_\text{sat})^n} 
\end{equation}
Assuming the emission probability distributions of the emitters are independent and identical, the total intensity becomes proportional to the probability of a single emitter being excited at least once.
\begin{subequations}
\begin{align}
    I(P) &= A(1-\exp(-(P/P_\text{sat})^n),\label{eq:pl_power_dependence}\\ &\simeq A(P/P_\text{sat})^n, \quad \text{for\ } P\ll P_\text{sat},
\label{eq:pl_power_dependence_low}\\
    &\simeq A, \quad \text{for\ } P\gg P_\text{sat},
    \label{eq:pl_power_dependence_high}
\end{align}    
\end{subequations}
where we have introduced the proportionality constant $A$ that maps from emission probability to emitted power. We see that the expression reduces to the familiar power law in the case of low powers. Any losses in the experimental setup is also accounted for by the factor $A$. We did a power scan on two samples annealed at \SI{360}{\celsius} and \SI{525}{\celsius} respectively, using the 405 nm laser. The brightness of the  centreline can be seen in Fig. \ref{fig:power_dependce} as a function of input power. The beam size here was estimated to be $\text{area}=\pi r^2 , r\simeq 100\ $\textmu m.

\subsection{Excitation Wavelength Dependence of Power Deposition}
\label{sec:supp:powerDeposition}

\begin{figure}[H]
    \centering
    \includegraphics[width=\linewidth]{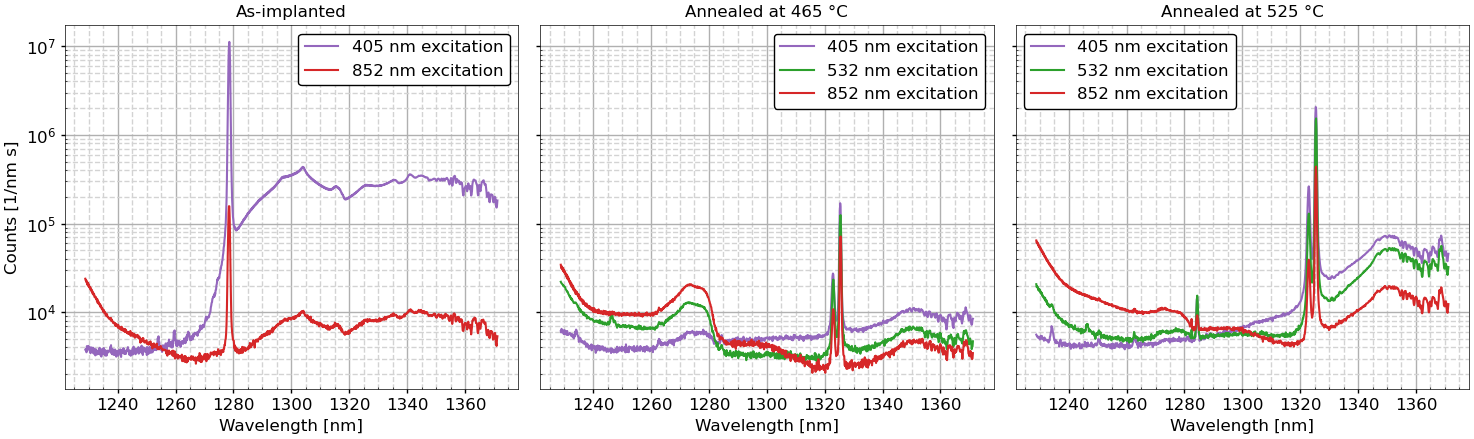}
    \caption{Plots of the same spectral window for different excitation laser on different samples. The left plot is from a sample with no implantation. The central plot is for a sample that has been annealed at \SI{465}{\celsius} and the right plot is for sample that has been annealed at \SI{525}{\celsius}. There is a clear trend that lower wavelengths gives a brighter emission spectrum for the central the colour centres while the material response is less systematic.}
    \label{fig:colour_scan_spectra}
\end{figure}

\begin{figure}[H]
    \centering
    \includegraphics[width=1\linewidth]{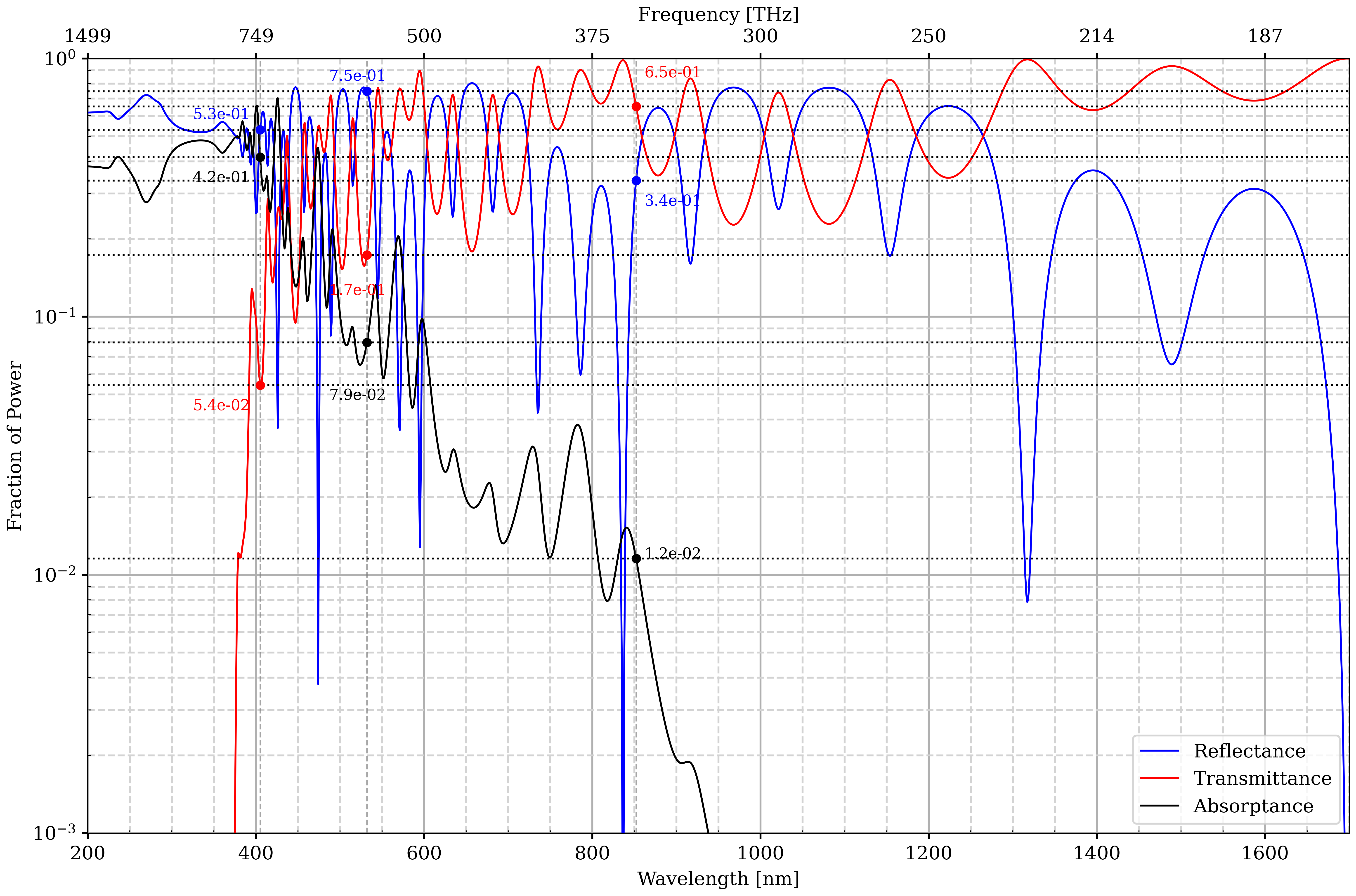}
        \caption{Reflectance, transmittance, and absorptance relative to the reference power for an incident angle of 30° and polarization parallel to the plane of incidence. For wavelengths below 400 nm, most of the light is absorbed during the first pass of the membrane. Above 400 nm, interference effects occur between reflections at the first and second interfaces. Overall, shorter wavelengths deposit more energy in the 220 nm-thick silicon membrane. Below 400 nm, increased reflectance from the first interface is observed due to the higher refractive index of silicon. The absorptance is near maximum at the 405 nm laser wavelength where ~30 to 50 \% of the power is absorbed in the material. To improve the model, the plane-wave assumption should be relaxed to account for the finite spot size. Furthermore, the material properties used in this model assume pure silicon and therefore is supected to differ from the implanted material.}
    \label{fig:absorptance_wavelength}
\end{figure}

\begin{figure}[h]
    \centering
    \includegraphics[width=1\linewidth]{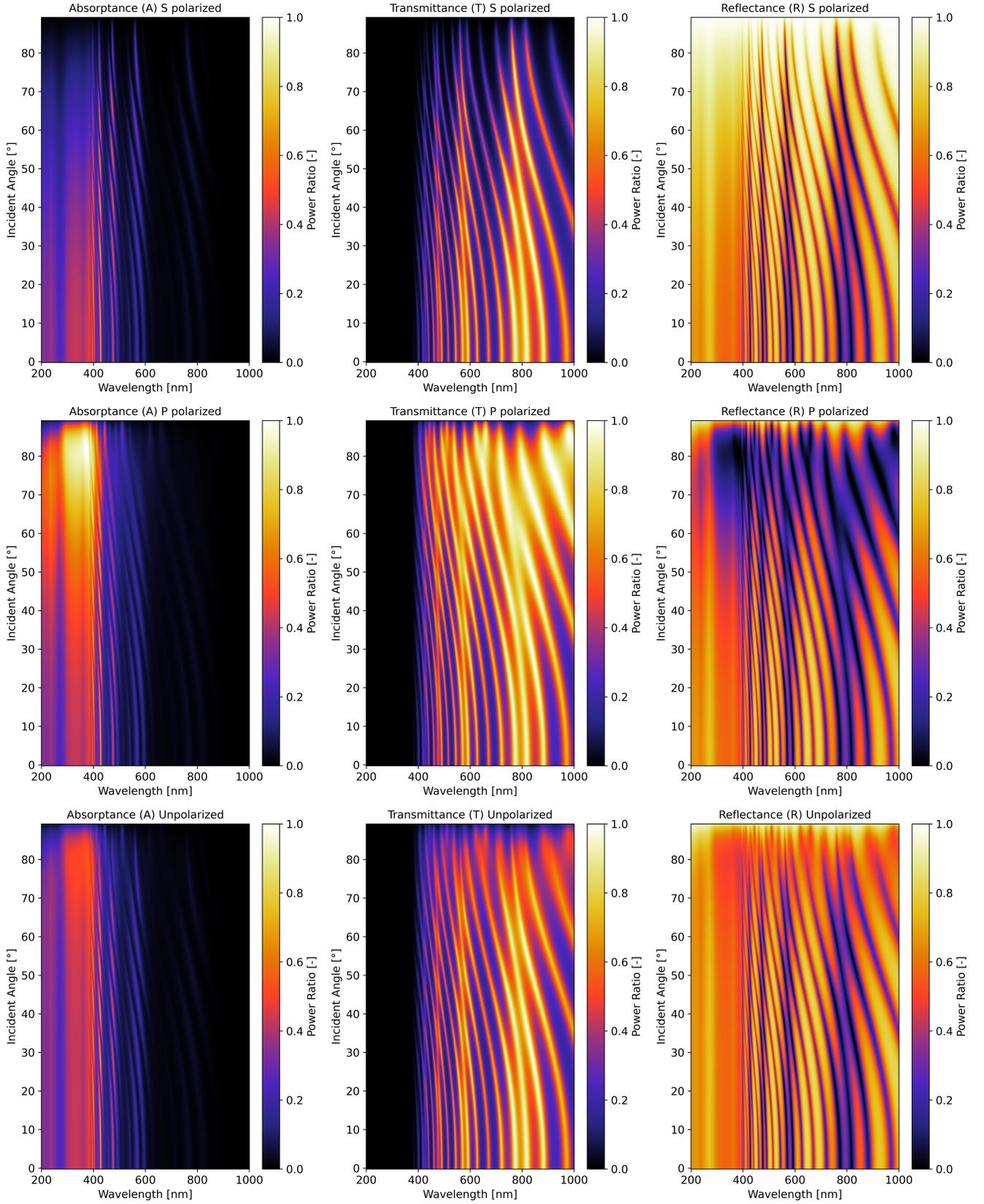}
    \caption{Extensive simulations of the absorptance, transmittance, and reflectance in the top layer of the SOI stack were performed using the transmission matrix method at different wavelengths, angles, and polarizations.}
    \label{fig:absorptance_simulation}
\end{figure}

The brightness of the colour centres was found to depend greatly on the excitation wavelength -- see Fig.~\ref{fig:colour_scan_spectra} and Tab.~\ref{tab:wavelength_dependence}. To understand the dependence on excitation wavelength we modelled the expected absorptance, transmittance, and reflectance of our stack at different free space wavelengths, angles and polarisations. The absorptance refers only to absorption of energy in the 220 nm thick silicon membrane assuming a 700 nm box layer and an infinite (in this case a reasonable approximation for 3 mm) silicon substrate.
A model for the wavelengths dependence of refractive indices is described in \cite{FrantaSilica2016} for silica and \cite{FrantaSilicon10K2017} for silicon. The simulation assumes that the input wave is a plane wave. This assumption breaks down for multiple reflections where the limited spots size might not overlap with itself from previously when coming in a an angle. Finally, the simulation assumes perfect interfaces with no roughness. This is justified since the interface roughness, of the order of 1 nm, is much smaller than the characteristic size of the light. The simulation was done in python using the TMM package that uses the transmission matrix method. Results of this are shown in Fig.~\ref{fig:absorptance_wavelength}. From the simulation we clearly see a greater absorptance below wavelengths of 400 nm. This corresponds to the skin depth of silicon being comparable to the membrane thickness of 220 nm. For a full study of the parameter space at different wavelengths, angles and polarizations see Fig.~\ref{fig:absorptance_simulation}.

\begin{table}[H]
    \centering
    \setlength{\tabcolsep}{12pt}
    \begin{tabular}{lrrrr}
    \toprule
     Excitation & TX0 Relative Area & TX1 Relative Area & G centre Relative Area & W centre Relative Area \\
    \midrule
    405 nm  & 1.00  & 1.00  & 1.00   & 1.00   \\
    532 nm  & 0.702 & 0.476 & NaN    & 0.284  \\
    852 nm  & 0.202 & 0.130 & 0.0146 & NaN    \\
    \bottomrule
    \end{tabular}
    \caption{Table showing fitted relative peak heights relative to their own height when excited with 405nm as a function of the excitation used.}
    \label{tab:wavelength_dependence}
\end{table}

\section{Analysis Methods of the Photoluminescence Spectra}
\label{sec:supp:PLAnalysis}

To determine the central frequency, the amplitude, the width, and the area of the peaks in the PL spectrum, we fit the peaks of the W, G, C, I, and M centres to the sum of a Gaussian and a quadratic ''background'', which describes the SOI material response from the above band excitation. Hence, the counts, C, as a function of wavelength, $\lambda$ is given as
\begin{align}
    C(\lambda) = \frac{A}{\sigma \sqrt{2\pi}}e^{- \frac{(\lambda - \lambda_c)^2}{2\sigma^2}} + b_0 + b_1\lambda + b_2\lambda^2
\end{align}
where A is the area of the peak, $\lambda_C$ is the central frequency, $\sigma$ is the width, and $b_0$, $b_1$, $b_2$ represent the material response. For the TX0-TX1 double peak, we fit with a sum of two Gaussians and a quadratic background.\\
\\
\clearpage

The samples were measured across multiple cooldowns. For each cooldown, a sample annealed at \SI{465}{\celsius} was used as a reference to allow consistent comparison of PL intensities across different cooldowns. The TX0 peak in the spectrum recorded of the \SI{465}{\celsius} sample in the first cooldown was identified as the ''master'' reference, $C_\text{master}(1326 \text{ nm})$. To compensate for variations in excitation/collection efficiencies between cooldowns, all other spectra are normalised to the ratio of the reference sample that was part of the same cooldown, $C_\text{ref}(1326 \text{ nm})$. Hence, the raw brightness, $C_\text{raw}(\lambda)$, is converted to a normalised brightness, $C_\text{norm}(\lambda)$.
\begin{align}
    C_\text{norm}(\lambda) = C_\text{raw}(\lambda) \frac{C_\text{master}(1326 \text{ nm})}{C_\text{ref}(1326 \text{ nm})}
\end{align}

\section{Temperature Dependence of T centre PL Signal}
\label{sec:supp:tempRedShift}
\begin{figure}
    \centering
    \includegraphics[width=1\linewidth]{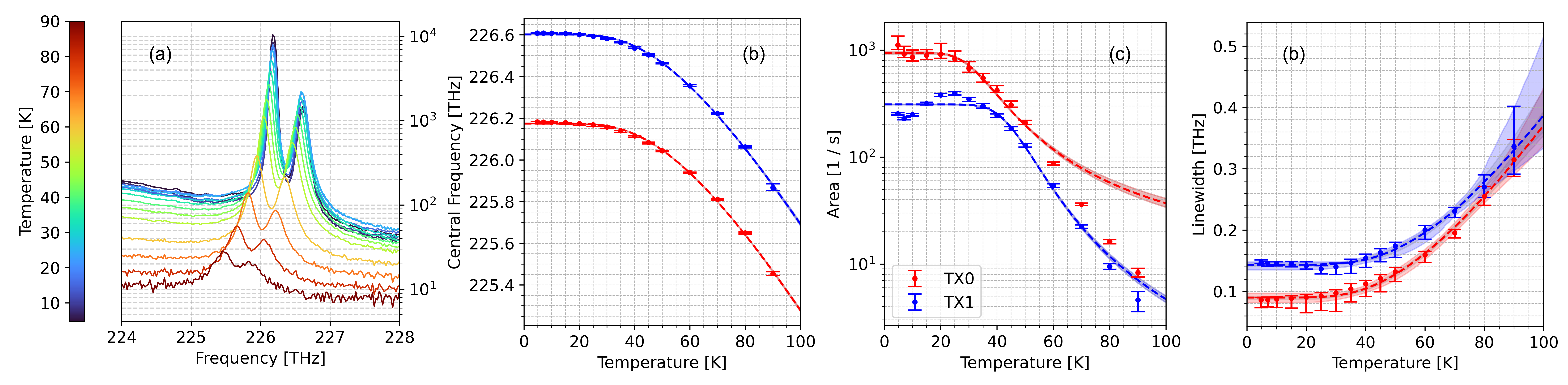}
    \caption{Temperature dependence of the T centre emission spectrum. \textbf{(a)} shows the spectrum of the T centre lines at different temperatures ranging from 4.8 K to 90 K. \textbf{(b-d)} Plots of the parameters of the double Gaussian fit to TX0 (red) and TX1 (blue) in the spectra in (a). \textbf{(b)} The area of the Gaussian fitted using Eq. \ref{eq:temperature_intensity_simple}. \textbf{(c)} The central frequency fitted using Eq. \ref{eq:temp_frequency_bose_einstein}. \textbf{(d)} The full width at half maximum fitted using Eq. \ref{eq:temp_fwhm}.}
    \label{fig:temperature_scan}
\end{figure}
We scan the temperature of the cryostat and sample from 4.8 K to 90 K; the result of which can be seen in Fig. \ref{fig:temperature_scan}(a). Increasing the temperature red-shifts the band gap of silicon as a result of increased phonon interactions in the crystal lattice, which in turn red-shifts the emission of the T centre (see Fig. \ref{fig:temperature_scan}(b)). A model, which has been shown to describe the temperature dependence of the band gap of silicon well, is the Bose-Einstein model \cite{Pässler1999}
\begin{equation}
    E(T) = E(0) - \frac{\alpha\Theta}{2}\left(\coth\left(\frac{\Theta}{2T}\right)-1\right)\label{eq:temp_frequency_bose_einstein},
\end{equation}
where $\alpha$ is a coupling constant that describes the strength of the interaction with the phonons, and $\Theta$ is a material dependent effective phonon temperature.\\

The emitter brightness decreases at higher temperatures, which can be attributed to the increased probability of the excited carriers populating states that decay non-radiatively, and hence do not contribute to the emitter signal. The simplest model, assuming only a single non radiative decay channel, is given by
\begin{equation}
    I(T) = \frac{I(0)}{1+ A \exp(-\frac{E_B}{k_BT})}.\label{eq:temperature_intensity_simple}
\end{equation}
where $E_B$ is the energy of the radiative decay channel. The fit to the extracted PL signals is seen in \ref{fig:temperature_scan}(c). At higher temperatures, the model breaks down and overshoots the brightness for both the TX0- and TX1-line, which is thought to be due to multiple non-radiative channels becoming significant \cite{Irion_1985, Sarihan2025}. \\

Finally, the linewidth increases at higher temperatures. We make use of a simple model for thermally activated homogeneous broadening \cite{simmons2018} given by
\begin{equation}
    \text{FWHM}= P_0 +\frac{P_T}{\exp(E_A/k_BT)-1},
    \label{eq:temp_fwhm}
\end{equation}
where $E_A$ is the energy at which the broadening become significant. The fit can be seen in Fig.~\ref{fig:temperature_scan}(d) where we get a transition energy of $E_A=16(14, 18)$ meV for the TX0-line and $E_A=19(16, 23)$ meV for the TX1-line. The model suggests that the broadening is caused by the phononic environment exciting higher order electronic states, separated by $E_A$ from the TX0 and TX1 lines, as these states allow decay through different channels.

\section{Detailed List of Silicon Colour Centres}
\label{sec:supp:detailedListSiCC}

\begin{figure}
    \centering
    \includegraphics[width=1\linewidth]{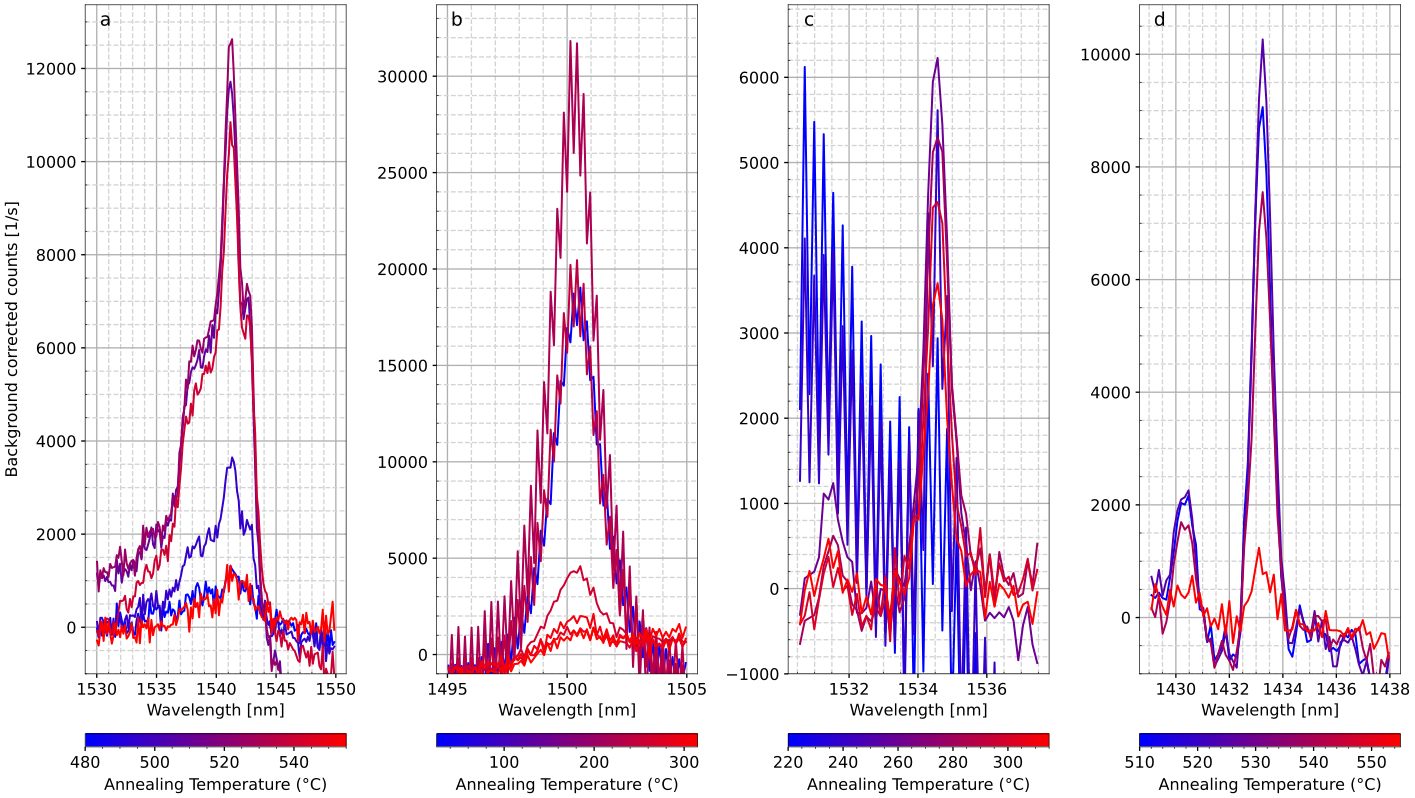}
    \caption{Spectra for a range of unidentified emitters.
    Panels \textbf{(a-d)} correspond to the lines listed in Table \ref{tab:silicon_defects_extended} as "Unknown1-4". The characteristic shape and very broad FWHM of Unknown1 might indicate a different and broader emission process than the other peaks, which are closer in shape to emission lines from other colour centres. Characterised photoluminescence background has been subtracted from the signal for ease of visualisation.}
    \label{fig:other_weirdpeaks}
\end{figure}

Table \ref{tab:silicon_defects_extended} shows an extensive list of emission lines of colour centres in silicon, their central wavelength, energy, including their general structure, and spin properties. It also includes associated lines such as the phononic transverse acoustic (TA) and transverse optical (TO) lines as well as local vibration modes (LVM). We also present 4 stable signals in our PL spectra which have yet to be explained in Fig.~\ref{fig:other_weirdpeaks}

\begin{table}[htbp]
    \centering
    \caption{Emission lines and properties of different colour centres in silicon.}
    \label{tab:silicon_defects_extended}
    \begin{adjustbox}{max width=1\textwidth, scale=1}
    \begin{tabular}{lcccccccccccccc}
        \toprule
        Name & ZPL [eV] & ZPL [THz] & ZPL [nm] & TA [eV] & TA [THz] & TA [nm] & TO [eV] & TO [THz] & TO [nm] & $\text{LVM}_1$ [eV] & $\text{LVM}_1$ [THz] & $\text{LVM}_1$ [nm] & Structure & Ref. \\
        \midrule
        X   & 1.04 & 251.5 & 1192.2 & -- & -- & -- & -- & -- & -- & -- & -- & -- & I\textsubscript{4} (quad-interstitial Si) & \cite{Hayama_2004} \\
        W   & 1.018 & 246.2 & 1217.9 & 1.0 & 241.6 & 1241.1 & 1.0 & 232.1 & 1291.5 & 0.9 & 229.3 & 1307.2 & I\textsubscript{3} (tri-interstitial Si) & \cite{Hayama_2004, Nikolskaya_2024} \\
        G   & 0.970 & 234.5 & 1278.2 & 1.0 & 230.0 & 1303.7 & 0.9 & 220.5 & 1359.5 & 0.9 & 216.9 & 1382.2 & C\textsubscript{s}-Si\textsubscript{i}-C\textsubscript{s} & \cite{Quard_2024, Beaufils_2018} \\
        I   & 0.965 & 233.3 & 1284.8 & 0.9 & 228.7 & 1310.6 & 0.9 & 219.3 & 1367.0 & -- & -- & -- & C-C-H(O) (unknown structure) & \cite{Gower_1997, Lightowlers_1994, SAFONOV1999} \\
        T & 0.935 & 226.1 & 1326.0 & 0.9 & 221.5 & 1353.5 & 0.9 & 212.1 & 1413.7 & 0.9 & 210.1 & 1426.9 & (C-C)\textsubscript{s}-H\textsubscript{i} & \cite{Clear_2024, macquarrie_generating_2021} \\
        H   & 0.926 & 223.9 & 1338.9 & 0.9 & 219.3 & 1367.0 & -- & -- & -- & -- & -- & -- & C-O (unknown structure) & \cite{Magnea_1984} \\
        Ci  & 0.860 & 206.8 & 1448.1 & -- & -- & -- & -- & -- & -- & -- & -- & -- & (Si-C)\textsubscript{Si} & \cite{Jhuria_2024} \\
        C   & 0.790 & 191.0 & 1569.4 & 0.8 & 186.4 & 1608.1 & 0.7 & 177.0 & 1693.8 & -- & -- & -- & C\textsubscript{i}-O\textsubscript{i} & \cite{Nakamura_1995, Silkinis_2025} \\
        P   & 0.767 & 185.5 & 1616.5 & -- & -- & -- & -- & -- & -- & -- & -- & -- & C\textsubscript{i}-O\textsubscript{2i} & \cite{Kurner_1989, Wagner_1985} \\
        M   & 0.761 & 184.1 & 1628.6 & -- & -- & -- & -- & -- & -- & -- & -- & -- & C-C-H (unknown structure) & \cite{Safonov_1997, Kuganathan_2025} \\
        CN   & 0.828 & 200.2 & 1497.4 & -- & -- & -- & -- & -- & -- & -- & -- & -- & (C-N)$_si$ (unknown structure) & \cite{nangoi2025cncomplexalternativet} \\
        D1  & 0.812 & 196.3 & 1526.9 & -- & -- & -- & -- & -- & -- & -- & -- & -- &
        Dislocation & \cite{Sauer1985-dt} \\
        D2  & 0.875 & 211.6 & 1417.0 & -- & -- & -- & -- & -- & -- & -- & -- & -- & Dislocation & \cite{Sauer1985-dt} \\
        D3  & 0.934 & 225.8 & 1327.5 & -- & -- & -- & -- & -- & -- & -- & -- & -- & Dislocation & \cite{Sauer1985-dt} \\
        D4  & 1.000 & 241.8 & 1239.8 & -- & -- & -- & -- & -- & -- & -- & -- & -- & Dislocation & \cite{Sauer1985-dt} \\
        \midrule
        & & & & & & & & Unidentified emitters & & & & & & \\
        Name & ZPL [eV] & ZPL [THz] & ZPL [nm] & TA [eV] & TA [THz] & TA [nm] & TO [eV] & TO [THz] & TO [nm] & $\text{LVM}_1$ [eV] & $\text{LVM}_1$ [THz] & $\text{LVM}_1$ [nm] & Structure & Temp range \\
        \midrule
        CN*   & 0.828 & 200.3 & 1496.7 & -- & -- & -- & -- & -- & -- & -- & -- & -- & -- &  520-600 $^\circ$C\\
        Unknown1 & 0.804 & 194.4 & 1542.0 & -- & -- & -- & -- & -- & -- & -- & -- & -- & -- & 480-555 $^\circ$C\\
        Unknown2 & 0.826 & 199.8 & 1500.5 & -- & -- & -- & -- & -- & -- & -- & -- & -- & -- & implanted-315 $^\circ$C \\
        Unknown3 & 0.808 & 195.4 & 1534.5 & -- & -- & -- & -- & -- & -- & -- & -- & -- & -- & 220-315 $^\circ$C  \\
        Unknown4 & 0.865 & 209.1 & 1433.5 & -- & -- & -- & -- & -- & -- & -- & -- & -- & -- & 510-555 $^\circ$C  \\

        \bottomrule
    \end{tabular}
    \end{adjustbox}
\end{table}

\FloatBarrier

\section{Effects of Boiling, Gas Flow, and Thermal Control}
\label{sec:supp:qualitative_variations}

\begin{figure}
     \includegraphics[width=0.48\linewidth]{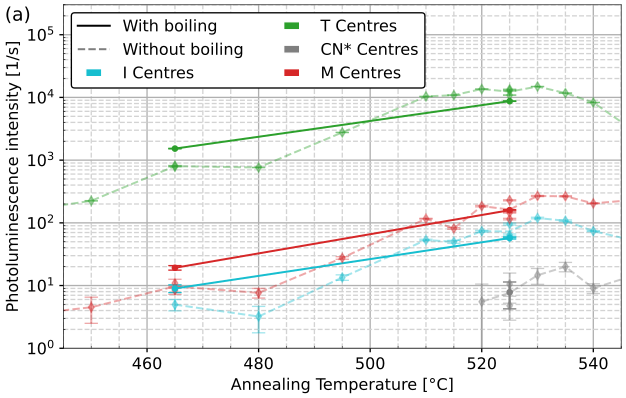}
     \includegraphics[width=0.48\textwidth]{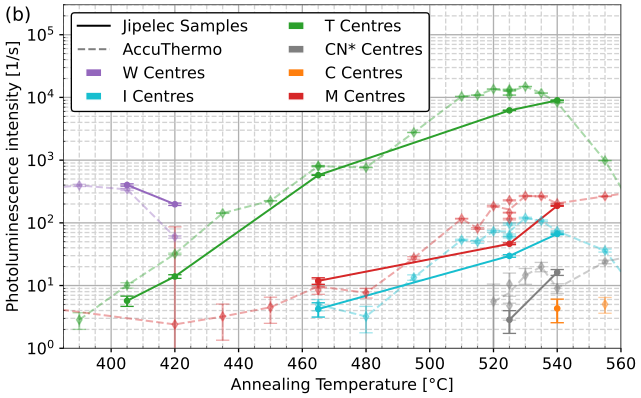}
    \caption{\textbf{a)}: Shows our results from samples that were boiled for 1 hour in deionized water immediately before the activation anneal was performed, showing no clear difference in optimal temperature or significant deviation from the control samples. \textbf{b)}: Shows samples annealed in a different tool, where temperature was controlled with an optical pyrometer. The optimal temperature seems to have shifted slightly up to $>$\SI{525}{\celsius}, perhaps indicating that our primary rapid thermal processor was underestimating the sample temperature.}
    \label{fig:Boiling_Jipelec}
\end{figure}

Previous studies \cite{Dhaliah22} have suggested that the partial pressure of hydrogen during the activation anneal might be crucial for generating a high yield of T centres due to the over and under-hydrogenated versions. In particular, boiling samples in deionized water has been presented as a method of increasing the efficiency of T centre generation \cite{macquarrie_generating_2021}. To investigate whether our process for T centre generation could be further optimized and whether this might explain the difference of optimal temperature we found compared with the existing literature, we boiled two samples for 1 hour in deionised water before annealing them at 465 and \SI{525}{\celsius}. The results of the following PL measurements are presented in Figure~\ref{fig:Boiling_Jipelec}a and show no significant impact of the boiling on the density of T centres. \\

\begin{figure}
    \centering
    \includegraphics[width=0.5\linewidth]{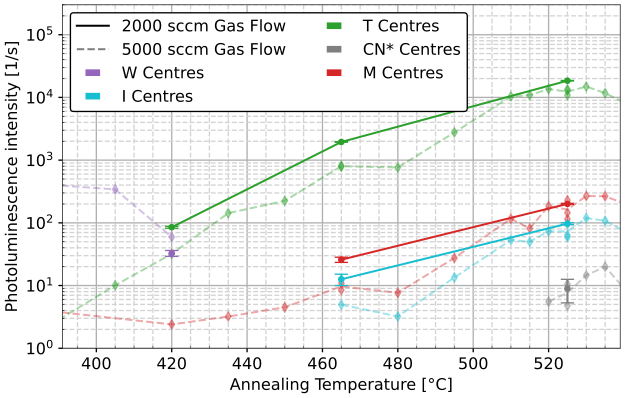}
    \caption{Results from reducing the flow of N$_2$ into the process chamber to 2000 sccm (solid lines) instead of the 5000 sccms (dashed lines) used in the main text. We see no effect from the difference in gas flow or from ambient hydrogen in the PL signals.}
    \label{fig:reduced_gas_flow}
\end{figure}

In the oven used to anneal the samples presented in the main paper, the temperature was measured with a thermocouple attached to the susceptor containing the chip. A high flow of 5000 sccm of nitrogen gas through the oven ensures a pure atmosphere of nitrogen, but this might contribute to local cooling of the sample, leading us to overestimate the actual temperature of the sample. To investigate this, 5 samples were annealed in a different RTP tool (Jipelec JetFirst 200 RTP) which had a vacuum chamber and turbo pump allowing us to run the process with a much lower flow of process gas (200 sccm) without risking contamination from the ambient air. The temperature in this tool was measured by an optical pyrometer and the sample placed in an unlidded susceptor on a carrier wafer, which while less accurate, is not vulnerable to the same systematic bias as a thermocouple. The results of the runs on the Jipelec tool are presented in Figure~\ref{fig:Boiling_Jipelec}b where we see a slightly higher optimal annealing temperature, which is consistent with our thermocouple in our primary RTP tool being placed on the bottom of the susceptor farthest from the heat source, probably leading to a slight underestimation of the actual sample temperature when measured with the thermocouple. \\
\\
We also investigated whether a lower gas flow of 2000 sccm would result in reduced cooling of the sample and therefore a lower optimal temperature as measured by our thermocouple, but our results are consistent with our previous trials as seen in Figure~\ref{fig:reduced_gas_flow}.

\FloatBarrier
\section{Fabrication Integration} 
\label{sec:supp:fabrication}

As part of investigating formation and decomposition processes we also investigated compatibility of colour centres with common fabrication methods. Apart from the results presented in the main text, we exposed a series of chips to different steps of an e-beam lithography process. \\

\begin{figure}[h]
    \centering
    \includegraphics[width=1\linewidth]{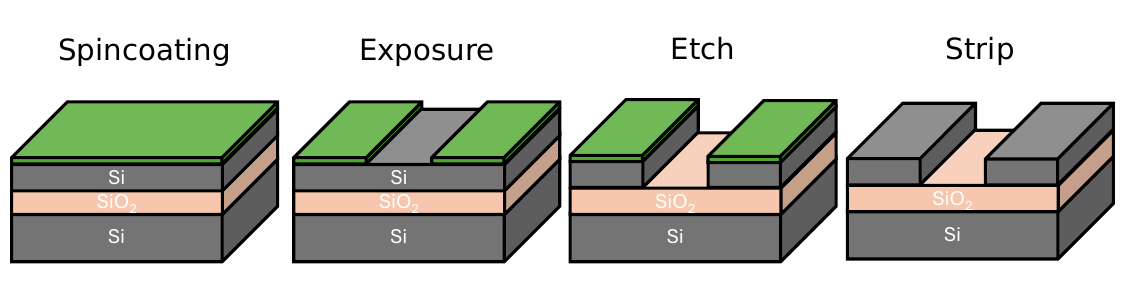}
    \caption{An illustration of the 4 main steps in making a photonic device using e-beam lithography}
    \label{fig:Fabrication_workflow}
\end{figure}

The applied fabrication process can be divided in four phases briefly illustrated in Fig~\ref{fig:Fabrication_workflow}. Samples are first spin-coated with $\approx 100$~nm CSAR (Allresist Ar-P 6200) which is soft-baked at \SI{185}{\celsius} for 2 min to evaporate the anisole contents. In the next phase, a device pattern is exposed in the e-beam resist (with a calculated dose of 300 \textmu C/cm$^2$) after which the exposed areas are dissolved in O-Xylene. The sample is then descummed in a remote oxygen-plasma for 45 s and hard-baked at \SI{110}{\celsius} for 2 minutes. In the third phase, the sample is etched in an SF$_6$/C$_4$CF$_8$ plasma at room temperature. The final phase consists of a 4 minute direct oxygen plasma to get rid of remaining resist and any fluorocarbons deposited by the Bosch process. \\ 

\begin{figure}
    \centering
    \includegraphics[width=0.6\linewidth]{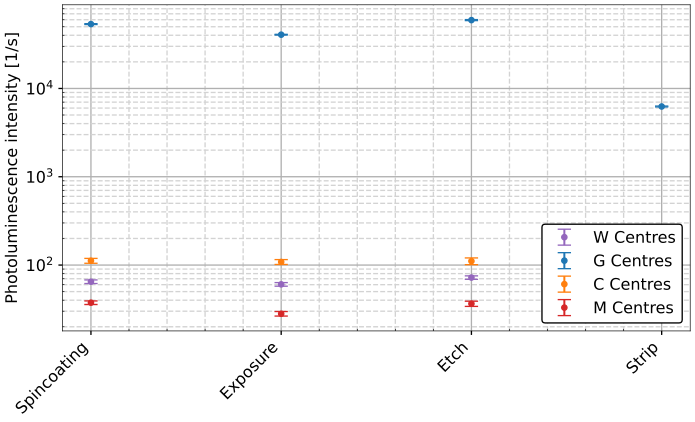}
    \caption{We see that the as-implanted colour centres remain present at each step of fabrication, only reducing in intensity noticeably at the end when oxygen plasma is used, and where the W, C, and M centres become too weak to resolve.}
    \label{fig:Fabrication_scan}
\end{figure}

A series of 4 samples were made, stopping after respectively the spin-coating, the exposure, the development, and the resist strip phases as described above. The results of PL measurments on these samples can be seen in Figure~\ref{fig:Fabrication_scan}. They indicates that the exposure to direct oxygen plasma is by far the harshest step in the process, leading to degradation of the G centres. \\

\begin{figure}
    \centering
    \includegraphics[width=1\linewidth]{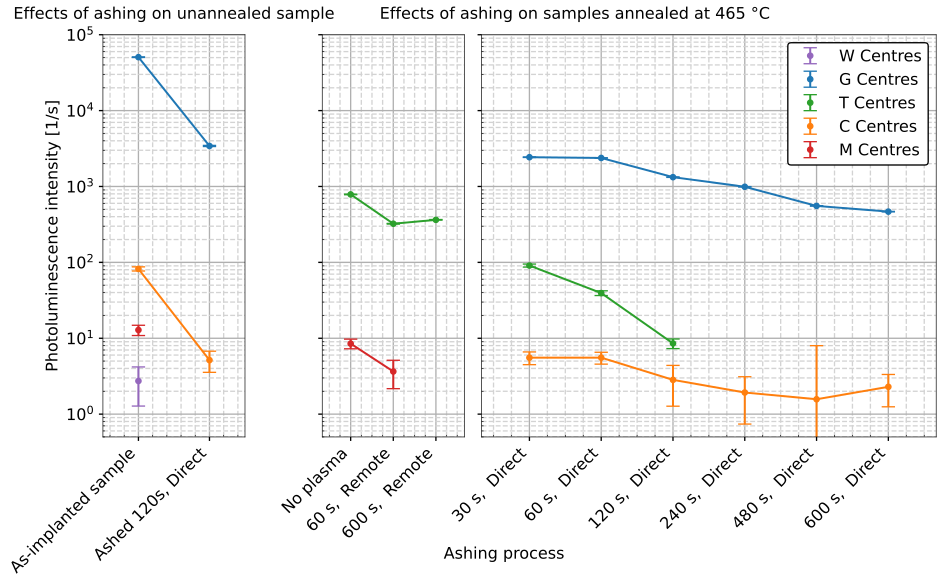}
    \caption{Comparison of ashing on a sample dependent on whether or not it has undergone the activation anneal. We see that ashing on its own tends to break down  centre and C centres which are present in the sample after implantation, but that the same ashing process on an annealed sample seems to "reactivate" centres which are otherwise annealed out at these temperatures. Also of note is that remote ashing seems to have a negligible effect on centre densities, leaving it as a viable alternative for sample cleaning in situations where extra care is required.}
    \label{fig:ashing_times}
\end{figure}

This laid the groundwork for the further investigations of the susceptibility of the T centre to decomposing during an oxygen plasma. In addition to the results shown in the main paper, we show the effect of ashing on G centres. As presented in the main paper, direct immersion of an annealed sample in the oxygen plasma has the effect of strongly increasing the density of G centres. However as shown in figure~\ref{fig:ashing_times} this is only true for samples where the G centres have previously been annealed out. A non-annealed sample will see a reduction in the G centre density from direct ashing, and similarly, additional ashing after the first 30 s tends to reduce the density of G centres. It is likely that the plasma is actually breaking up some non-optical defect states and decomposing them into G centres which are then themselves decomposed further by the ashing. \\

\end{document}